\documentclass[sigconf]{acmart}

\usepackage[toc,page]{appendix}
\usepackage{subfigure}
\usepackage{multirow}
\usepackage{wrapfig}
\usepackage{soul}
\setcopyright{none}

\begin{document}
\title{Privacy-Preserving and Efficient Verification of the Outcome in Genome-Wide Association Studies}
\titlenote{Published in the Proceedings on Privacy Enhancing Technologies (PoPETs), Vol. 2022, Issue 3, 2022.}

\author{Anisa Halimi}
\authornote{This research was undertaken while the author was at Case Western Reserve University.}
\email{anisa.halimi@ibm.com} 
\affiliation{
	\institution{IBM Research Europe - Dublin}
}
\author{Leonard Dervishi}
\email{leonard.dervishi@case.edu}
\affiliation{
	\institution{Case Western Reserve University}
}
\author{Erman Ayday}
\email{erman.ayday@case.edu}
\affiliation{
	\institution{Case Western Reserve University}
}
\author{Apostolos Pyrgelis}
\email{apostolos.pyrgelis@epfl.ch}
\affiliation{
	\institution{EPFL}
}
\author{Juan Ram\'on Troncoso-Pastoriza}
\email{juan.troncoso-pastoriza@epfl.ch}
\affiliation{
	\institution{EPFL}
}
\author{Jean-Pierre Hubaux}
\email{jean-pierre.hubaux@epfl.ch}
\affiliation{
	\institution{EPFL}
}
\author{Xiaoqian Jiang}
\email{xiaoqian.jiang@uth.tmc.edu}
\affiliation{
	\institution{University of Texas, Health Science Center}
}
\author{Jaideep Vaidya}
\email{jsvaidya@business.rutgers.edu}
\affiliation{
	\institution{Rutgers University}
}

\begin{abstract}
Providing provenance in scientific workflows is essential for reproducibility and auditability purposes. Workflow systems model and record provenance describing the steps performed to obtain the final results of a computation. In this work, we propose a framework that verifies the correctness of the aggregate statistics obtained as a result of a genome-wide association study (GWAS) conducted by a researcher while protecting individuals' privacy in the researcher's dataset. In GWAS, the goal of the researcher is to identify highly associated point mutations (variants) with a given phenotype. The researcher publishes the workflow of the conducted study, its output, and associated metadata. In the proposed framework, the researcher keeps the research dataset private while providing, as part of the metadata, a partial noisy dataset (that achieves local differential privacy). To check the correctness of the workflow output, a verifier makes use of the workflow, its metadata, and results of another GWAS (conducted using publicly available datasets) to distinguish between correct statistics and incorrect ones. For evaluation, we use real genomic data and show that the correctness of the workflow output (i.e., that the output is computed correctly by the researcher) can be verified with high accuracy even when the aggregate statistics of a small number of variants are provided. We also quantify the privacy leakage due to the provided workflow and its associated metadata and show that the additional privacy risk due to the provided metadata does not increase the existing privacy risk due to sharing of the research results. Thus, our results show that the workflow output (i.e., research results) can be verified with high confidence in a privacy-preserving way. We believe that this work will be a valuable step towards providing provenance in a privacy-preserving way while providing guarantees to the users about the correctness of the results.
\end{abstract}
\keywords{Privacy; verifiable computation; genome-wide association studies; scientific workflows; provenance} 

\maketitle

\section{Introduction}

Provenance refers to the place of origin or earliest known history of an object in question. In the modern era, especially in scientific research, it is becoming crucial to provide provenance to allow reproducibility and auditability. In most applications, provenance is captured through workflows~\cite{myexperiment,oinn2004taverna,bowers2005actor,handigol2012debugger}. For researchers, it is crucial to verify the correctness of a published research, especially if they are planning to use the research findings in their study. Computational errors might occur during the workflow (e.g., the published results/statistics or the metadata may be computed wrong) or during the quality control~\cite{zuvich2011pitfalls,turner2011quality} (e.g., a researcher might use low quality data to conduct the research). It is trivial to verify the correctness of the research findings if, besides the workflow and its associated metadata, the input dataset is provided. However, the input dataset might not always be released as it may contain sensitive information about individuals (e.g., personal records in the dataset). In such cases, verifying the correctness of the computations becomes non-trivial. There exist several works in the field of verifiable computation~\cite{walfish2015verifying,yu2017survey}, which aim to do various computations on the cloud while verifying the correctness of the returned results. However, the problem we consider cannot be directly solved using existing verifiable computation techniques because in our case, a researcher aims to verify the correctness of a published research without having access to the dataset. One alternative may be to use ``homomorphic authenticators''~\cite{gennaro2013fully}, but they are impractical for statistical analysis on large datasets due to the high computation.

In this work, we propose a framework that efficiently verifies the correctness of the statistics obtained as an outcome of a genome-wide association study (GWAS) while preserving the privacy of the research dataset (i.e., individuals' personal data in the dataset). GWAS is a popular method for identifying genetic variations (mutations) that are associated with a particular phenotype (disease). For researchers, showing that discovered associations are correctly computed and the results are reproducible is of immense importance. At the same time, GWAS studies include highly sensitive datasets that contain genomic and phenotypic information of individuals that participate in the study. Genomic data includes privacy-sensitive information about an individual, such as ethnicity, kinship, and predisposition to certain diseases, while phenotype data may include their diagnosis (disease status). Due to such privacy concerns, the research dataset is typically kept private. We consider a scenario between (i) a researcher, which conducts GWAS and publishes the research findings (i.e., $p$-value, odds ratio, or minor allele frequency) along with the workflow and its associated metadata and (ii) a verifier, which receives the research results and is willing to check their correctness. The metadata includes the trait (phenotype) being studied, the population of the users in the study, the number of users in the case and control groups, and the number of genetic variants (single nucleotide polymorphisms - SNPs). In addition to these, in the proposed framework, we also include (i) a partial noisy dataset (generated by the researcher to achieve local differential privacy - LDP) including only the identified genetic variants that have high associations with the studied phenotype and (ii) the distribution of the noise added to construct the partial dataset (e.g., privacy parameter of LDP) as a part of the metadata, which helps for the verification of the GWAS output. 

To check the correctness of the provided associations (i.e., published statistics by the researcher as a result of GWAS), first, the verifier performs the same statistical study as the one executed by the researcher on the received partial noisy dataset. Then, by using a publicly available dataset, the verifier compares (i) the deviation between the published statistics and the ones computed from the received partial noisy dataset to (ii) the deviation between the statistics computed on a public dataset and the ones computed from the noisy version of the same public dataset (the verifier locally computes the noisy version of the public dataset). We show that this comparison statistically provides a proof to the verifier about the correctness of the researcher's computation. We observe that if the researcher provides incorrect results (due to miscalculations) and/or makes errors during the generation of the provided metadata, the verifier's confidence about the incorrectness of such results increases as the researcher deviates more from the original results. As we show via simulations, if the researcher's published statistics deviate less from the original values, the verifier's confidence decreases, but such small deviations still allow the verifier to receive high-quality results. We also show that the proposed scheme is robust for the selection of the public datasets that are used for verification.

Previous works have shown that aggregate statistics published as part of GWAS are prone to membership inference attacks~\cite{homer2008resolving,wang2009learning,sankararaman2009genomic}. In membership inference attacks, the attacker aims to determine whether the data of a target victim is part of the research dataset or not. The power of the membership inference attacks depends on the number of published statistics. In spite of this known vulnerability, sharing such statistics is crucial for research and it is approved by many institutions, including the NIH~\cite{nih}. Thus, one of our goals is to guarantee that the privacy risk for the dataset participants due to the partial noisy dataset (provided as part of the metadata) does not go beyond the \textit{baseline risk} that occurs due to the shared statistics as a result of GWAS. 

For evaluation, we use real genome data from OpenSNP~\cite{opensnp} and 1000 Genomes project~\cite{1000genomes} datasets. We particularly evaluate the confidence of the verifier about the correctness of the research output when a researcher (unintentionally) oversells them (e.g., reports stronger associations than the original ones). This is essential because the research findings can be used on other research studies (e.g., personalized medicine). When we set the number of returned statistics to $100$, we show that the verifier can correctly classify up to $90$ out of $100$ correct statistics (i.e., $p$-value, odds ratio, or minor allele frequency) provided by the researcher. Also, the verifier can detect at least $90$ out of $100$ incorrect statistics when, for example, the returned $p$-values deviate by at least $0.045$ from their correct values (e.g., when the researcher attempts to oversell or undersell the research findings). For instance, if the researcher (as a result of GWAS) obtains a $p$-value of $.09$ (widely accepted as a weak association) for a SNP and (erroneously) reports a $p$-value of $.045$ (widely accepted as a strong association) for the same SNP, the verifier can detect this with a probability close to $1$. We also show that the proposed framework can correctly classify all the correct statistics that are highly associated with the considered phenotype and all the incorrect statistics that imply a significant overselling of the real outcome. Overall, our results show that verifier's confidence decreases only when the returned statistics' values are close to the correct ones, which still results in high utility for the provided statistics. We also observe that the shared GWAS statistics bound the privacy vulnerability (for the research participants) and that the provided metadata (as part of the proposed framework) does not further increase this vulnerability. Therefore, the proposed framework does not introduce an additional privacy vulnerability for the research participants.
\section{Related Work}\label{sec:related_work}

\noindent{\bf Privacy and Security in Provenance and Workflows.} 
Both provenance privacy~\cite{chebotko2008scientific,gil2007privacy,gil2010reasoning,davidson2011provenance} and security~\cite{braun2008securing,hasan2007introducing,lyle2010trusted} have been studied in workflows. However, these works focus specifically on access control. Chebotko et al.~\cite{chebotko2008scientific} propose a mechanism that provides a partial view of a scientific workflow respecting the access privileges on the workflow input/output and the connections between the modules. Davidson et al.~\cite{davidson2011provenance} identify three types of privacy concerns in scientific workflows: data, module, and structural privacy. They discuss these privacy concerns via an example, focusing on the users' access privileges. Managing the access privileges of each user role and ensuring that they use the workflow along with the metadata in an intended way is challenging. Different from these works, we consider a scenario, where the workflow and its associated metadata are publicly available. 

Another line of work focuses on provenance sanitization~\cite{biton2008querying,cohen2008addressing,chebotko2008scientific,blaustein2011surrogate,dey2011propub,davidsonprovenanceview,missier2013provenance,cadenhead2011transforming}. By provenance sanitization, researchers refer to the general problem of ensuring that provenance solutions satisfy disclosure, privacy, and security requirements. Provenance sanitization is achieved via provenance graph transformation, where the structure of a scientific workflow is modified to satisfy all the requirements. A workflow is defined as a directed graph capturing the steps of a (scientific) process. A provenance graph is a directed acyclic graph representing the execution of the workflow. There exist three main approaches for provenance sanitization: (i) hiding, which eliminates the sensitive graph components, potentially leading to dangling edges and nodes, (ii) grouping, where several nodes are combined into one aggregate node, and (iii) anonymization, where the sensitive attributes of the nodes or edges are removed. Anonymization provides a better utility (in terms of how similar the transformed graph is to the original graph) than grouping because it preserves the structure of the workflow. Cheney et al.~\cite{cheney2014analytical} provide a comprehensive review of seven approaches for provenance sanitization. Mohy et al.~\cite{mohy2016comprehensive} propose a workflow provenance sanitization approach, called ProvS, which combines both anonymization and grouping. As opposed to these works, we focus on the verification of the correctness of the workflow output in GWAS, which, to the best of our knowledge, has not been studied before. At the same time, we make sure that the vulnerability (in terms of privacy risk) of the dataset participants does not increase due to the shared workflow and its associated metadata. 

\noindent{\bf Verifiable Computation.} There exists several schemes that offer verifiability with different characteristics. Their goal is to verify the correctness of the computations outsourced to a cloud. Yu et al.~\cite{yu2017survey} provide a broad review of some of the existing work in verifiable computation. Gennaro et al.~\cite{gennaro2010non} present a scheme based on garbled circuits and fully-homomorphic encryption. Backes et al.~\cite{backes2013verifiable} propose the incorporation of homomorphic MAC to verify the correctness of the computation done on an untrusted server. Parno et al.~\cite{parno2013pinocchio} propose a system, called Pinocchio, that uses quadratic arithmetic programs combined with a highly efficient cryptographic protocol. Trinoccio~\cite{schoenmakers2015trinocchio} improves Pinocchio by providing input privacy. ADSNARK~\cite{backes2015adsnark} extends Pinocchio~\cite{parno2013pinocchio} by proving computations done on authenticated data in a privacy-preserving way. Costello et al.~\cite{costello2015geppetto} propose Geppetto, a system that aims to further reduce prover overhead and increase its flexibility. Fiore et al.~\cite{fiore2016hash} extend Geppetto~\cite{costello2015geppetto} by enabling the verifier to verify the proof against a commitment done on some inputs independent of the computation. However, either these techniques can not be used to solve the problem we consider or they are not practical to verify computations performed on large datasets.
\section{Background}\label{sec:background}

In this section, we briefly introduce the relevant genetic concepts, as well as local differential privacy.

\subsection{Genomic Background}\label{sec:genomic_background}
The human genome is encoded as a sequence of nucleotides with values in the set \{A, T, C, G\}. The whole-genome consists of 3 billion pairs of nucleotides where $99.9\%$ of the genome is identical between any two individuals and the remaining part is referred to as genetic variation. {\it Single nucleotide polymorphisms} (SNPs) is the most common genetic variation, which stems from differences in single nucleotides. In the vast majority of cases, a SNP is {\it biallelic}, i.e., it can take two possible {\it alleles} (nucleotides at a variant location).

Genome-wide association studies (GWAS) have become a popular method for identifying genetic variations that are associated to a particular trait or phenotype. The most common approach of GWAS studies is the case-control setup, where the genomes of the individuals that carry the trait or phenotype (cases) are compared with the genomes of the healthy individuals (controls). As a result, the study identifies the SNPs that are associated to a certain trait. Assuming that SNPs are biallelic, they take values 0, 1, and 2, representing the number of their minor alleles. Therefore, GWAS data for each SNP can be summarized as either a $3\times 2$ or a $2\times 2$ contingency table (as shown in Table~\ref{table:gwas}), where each cell shows the number of cases and control users having a particular value (0, 1, or 2) for a given SNP. For instance, $S_0$ denotes the number of case users having $0$ for a given SNP, $C_0$ denotes the number of control users having $0$ for a given SNP, and so on. The output of GWAS usually consists of the $p$-value, odds ratio, and minor allele frequencies (MAFs) for the most significant SNPs. The ability of GWAS to identify associations of genetic variations to a phenotype depends on the quality of the data~\cite{turner2011quality,zuvich2011pitfalls}. The usage of low quality data may lead to false associations. Thus, it is crucial to follow the quality control (QC) procedure. We briefly describe the details of the QC procedure in Appendix~\ref{app:qc_steps}. Although we mainly focus on verification of GWAS results due to computational errors, the proposed framework can also be used to verify the correctness of QC steps, as we discuss in Section~\ref{sec:qc}. 
\begin{table}[ht]
\centering
    \caption{A $3 \times 2$ contingency table (left) and a $2 \times 2$ contingency table (right).}
    \vspace{-5pt}
    \resizebox{0.232\textwidth}{!}{
    \begin{subtable}\centering
        \begin{tabular}{|lcccc|}
		\hline
		 & \multicolumn{3}{c}{Genotype} & \\ \cline{2-4}
		 & 0 & 1 & 2 & Total \\ \hline
		Case & $S_0$ & $S_1$ & $S_2$ & S \\
		Control & $C_0$ & $C_1$ & $C_2$ & C \\
		Total & $n_0$ & $n_1$ & $n_2$ & n \\ \hline
		\end{tabular}
    \end{subtable}
    }
    \hfill{}
    \resizebox{0.21\textwidth}{!}{
    \begin{subtable}\centering
        \begin{tabular}{|lccc|}
		\hline
		 & \multicolumn{2}{c}{Genotype} & \\ \cline{2-3}
		 & 0 & 1 & Total \\ \hline
		Case & $S_0$ & $S_{1,2}$ & S \\
		Control & $C_0$ & $C_{1,2}$ & C \\
		Total & $n_0$ & $n_{1,2}$ & n \\ \hline
		\end{tabular}
    \end{subtable}
    }
    \label{table:gwas}
    \vspace{-10pt}
\end{table}

\subsection{Local Differential Privacy}\label{sec:ldp} 
Local differential privacy (LDP)~\cite{duchi2013local,kairouz2014extremal} is a variation of traditional differential privacy~\cite{dwork2008differential} with additional restrictions. In this setting, there is no trusted third party, thus offering a stronger level of protection for users' data. Each user perturbs their own data before sharing it with a data aggregator, so the aggregator only observes the perturbed data. Formally, an algorithm $A$ satisfies $\epsilon$-local differential privacy ($\epsilon$-LDP) if and only if for any input $v_1$ and $v_2$:
$$Pr[A(v_1) = y] \le e^{\epsilon} Pr[A(v_2)= y],$$
for all $y \in Range(A)$, where $Range(A)$ denotes all the possible outputs of the algorithm $A$. 
The most common way of achieving $\epsilon$-LDP is the randomized response mechanism~\cite{warner1965randomized}. In randomized response, a user reports the true value of a single bit of information with probability $\textbf{p}$ and flips the true value with probability $1-\textbf{p}$, satisfying $(ln\frac{\textbf{p}}{1-\textbf{p}})$-LDP. A data aggregator collects the perturbed values from users and attempts to determine the frequencies of values (bits) among the population.

Wang et al.~\cite{wang2017locally} introduced a framework that generalizes several LDP protocols and proposed a fast and generic aggregation technique for frequency estimation. Here, we summarize direct encoding (DE) and integrate it in our proposed verification framework (in Section~\ref{sec:framework}). Assume that there are $n$ individuals that have values from the set $[d]=\{1, 2, \ldots, d\}$. The goal of the aggregator is to predict the number of individuals having a value $i$ ($i\in [d])$. In DE, there is no encoding of the values. For perturbation, each value is reported correctly with probability $\textbf{p}=\frac{e^{\epsilon}}{d - 1 +e^{\epsilon}}$ and is flipped to one of the remaining $d - 1$ values with probability $\textbf{q}=\frac{1}{d - 1 +e^{\epsilon}}$. The aggregator collects all perturbed values and estimates the value of $i$ ($i\in\{1, 2, \ldots, d\}$) as $\tilde{c}_i = \frac{c_i-n\textbf{q}}{\textbf{p} - \textbf{q}}$, where $c_i$ is the number of times $i$ is reported.
\section{System and Threat Models}\label{sec:system_threat_model}

In this section, we introduce our system and threat models.

\subsection{System Model}\label{sec:system_model}

\begin{figure}[ht!]
    \centering
    \includegraphics[scale=0.25]{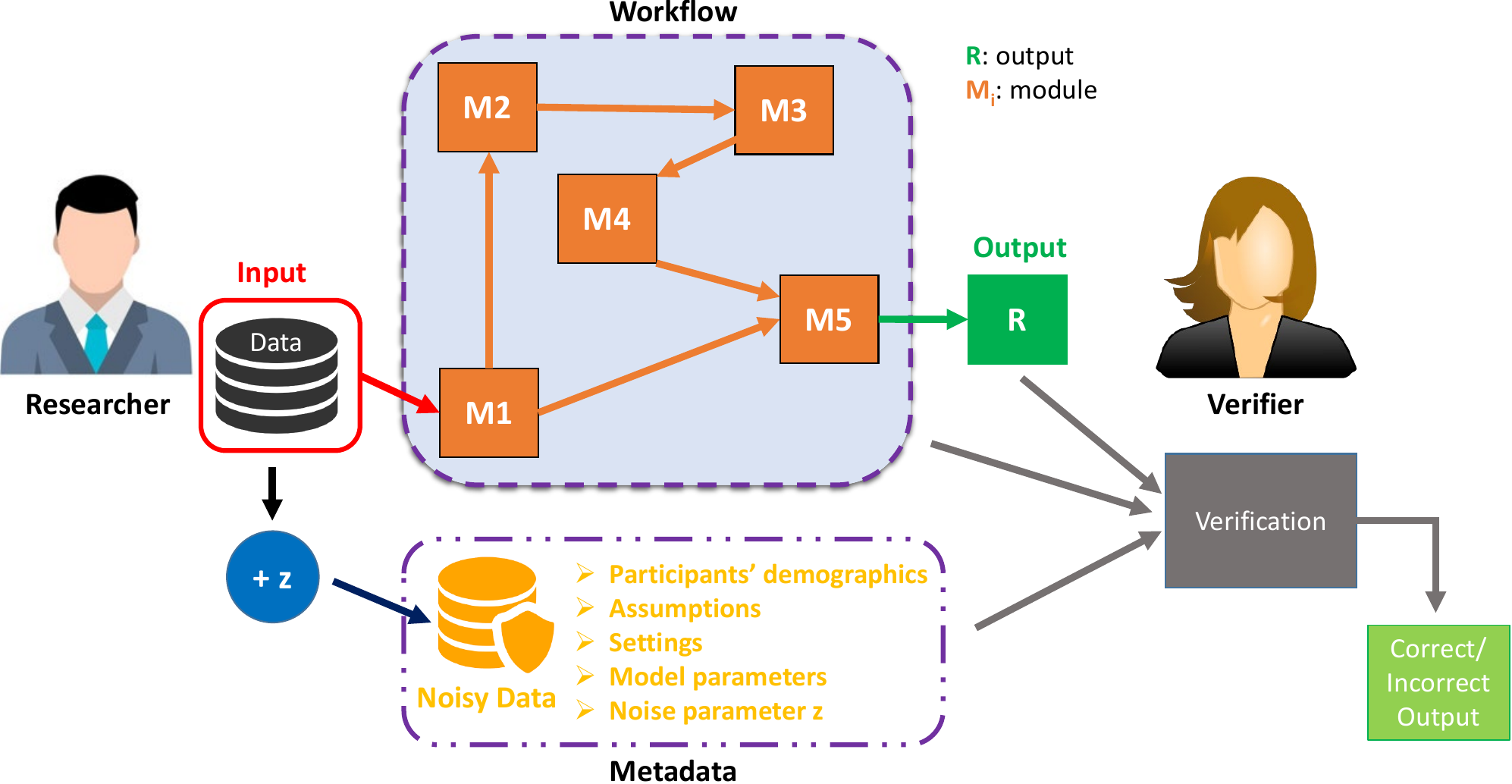}
    \caption{Overview of the proposed system model.}
    \label{fig:system}
    \vspace{-10pt}
\end{figure}

As shown in Figure~\ref{fig:system}, we consider a system that includes two parties: the researcher and the verifier. As discussed, we consider GWAS, in which the researcher's goal is to discover associations between the genetic variants (SNPs) and a type of trait/phenotype (e.g., cancer). For this, the researcher first constructs a dataset that includes genomic data of individuals that have cancer and healthy ones. The researcher creates a case-control setup, performs GWAS, and shares its workflow along with its associated metadata and the research findings. The research findings (workflow output) include the most highly associated SNPs and their corresponding statistics (e.g., $p$-value, odds ratio, and MAF). The workflow shows all the steps that are needed to perform GWAS. The metadata includes the phenotype being studied (cancer), the demographics of the research participants, the number of research participants, the number of SNPs, and a partial noisy dataset that includes the identified highly associated SNPs (as discussed in Section~\ref{sec:framework} the partial noisy dataset is used in the proposed verification algorithm). The researcher wants to ensure that the vulnerability of the dataset participants to genomic privacy attacks does not increase due to the provided metadata about the dataset. On the other hand, the verifier wants to check the correctness of the provided associations and use them in their research directly. Thus, it is crucial for the verifier to check the correctness of the results. The verifier has access to the workflow, its output, and metadata.  

\subsection{Threat Model}\label{sec:threat_model}

In the following, we summarize potential threats due to the researcher and the verifier.

\noindent\textbf{Researcher.} We assume an honest researcher that uses a legitimate dataset, rather than intentionally using a wrong/fake dataset. There is a huge incentive for the researcher to use a legitimate dataset (as will be discussed in Section~\ref{sec:limitations}). Our main goal is to develop a privacy-preserving and efficient verification tool for non-malicious researchers and research followers (i.e., verifiers), so that the researchers can have their results validated and adopted (by the verifiers or the research community) in a fast, privacy-preserving, and efficient way. Due to computational errors during GWAS or errors during the quality control (QC) steps~\cite{zuvich2011pitfalls,turner2011quality} (as described in Section~\ref{sec:genomic_background}), the researcher may unintentionally provide wrong results as the output of the research (GWAS). In this work we mostly focus on miscalculations done by the researcher during GWAS and discuss how to handle errors during QC in Section~\ref{sec:qc}. Due to such miscalculations, the researcher may unintentionally oversell (e.g., provide IDs of some weakly associated attributes along with strong statistics for such attributes) or undersell (e.g., provide the IDs of strongly associated attributes, but provide the statistics of some other attributes) the research results. Among these scenarios, the most harmful one is overselling the results due to miscalculations, since it may lead other researchers misuse reported strong associations in critical tasks (e,g., personalized medicine). Therefore, if the research reveals some new, strong associations with the considered trait, the verifier will most likely want to validate them before using in their research study. On the other hand, if the research does not reveal any new or strong associations, checking their correctness may not be equally interesting for the verifier. The verifier might be interested in checking such results only if they are also doing a similar study and believe that a stronger than the reported association should exist between a particular attribute and the studied trait. Thus, in this paper, we will mainly focus on a researcher that returns strong associations with the studied trait. However, as we will show in Section~\ref{sec:eval_verification}, our proposed framework can also verify the correctness of reported weak associations. Apart from doing computational errors during GWAS, the researcher may also compute the partial noisy dataset (a part of metadata that is used for verification, as discussed in Section~\ref{sec:system_model}) in a wrong way. Since we consider an honest (but error-prone) researcher, error due to miscalculations during GWAS and error in the generation of the partial noisy dataset are independent. We explore this in Section~\ref{sec:experimental_results}.

\noindent\textbf{Verifier.} The verifier may misbehave and try to infer sensitive information about the participants in the research dataset using the research output, metadata, and workflow. There exist known privacy attacks, such as membership inference~\cite{homer2008resolving,wang2009learning}, attribute inference~\cite{humbert2013addressing,deznabi2017inference}, and deanonymization attacks~\cite{gymrek2013identifying,humbert2015anonymizing} that exploit research results and/or partially provided datasets. In membership inference attacks, an attacker may attempt to determine whether a target record (victim) is part of the research dataset or not. In attribute inference attacks, the attacker aims at inferring additional private attribute(s) of an individual from the observed ones. For instance, a misbehaving verifier may try to infer hidden (unrevealed) genomic attributes (SNPs) from the SNPs in the partial noisy dataset. Here, the identities of the research participants are hidden, and hence attribute inference becomes a feasible attack scenario only after the attacker infers the membership of a victim to the research dataset. In deanonymization attacks, the goal of the attacker is to link the anonymized data (e.g., genome) of an individual to the individual's identity using some auxiliary information about the individual (such as observable phenotype or a part of the individual's genome). This may be possible due to the shared partial dataset as part of the metadata. However, in the proposed framework, the researcher shares only a small portion of the research dataset and applies noise to it before sharing. As a result, the most effective deanonymization attack can be conducted using the victim's partial genome as the auxiliary information (which makes deanonymization harder than membership inference attacks). Therefore, for our considered scenario, the most relevant attack for a misbehaving verifier is membership inference and we consider this attack in the rest of the paper. It has been shown that membership inference attacks against statistical biomedical datasets~\cite{homer2008resolving,wang2009learning,sankararaman2009genomic,backes2016membership} can be mitigated by adding noise to the released statistics to achieve differential privacy (DP)~\cite{uhlerop2013privacy,yu2014scalable,johnson2013privacy}. DP guarantees that the presence or absence of a data record does not significantly affect the released statistics. We discuss this extension and evaluate the performance of the proposed verification framework for GWAS statistics that are shared under DP in Section~\ref{sec:dp}. On the other hand, even though differentially private mechanisms can protect users against these attacks, they also come with a significant utility reduction (as also shown in Section~\ref{sec:dp}).
\section{Proposed Framework}\label{sec:framework}

In this section, we first introduce the GWAS statistics we consider. Then, we describe how the proposed framework works for privacy-preserving verification of GWAS results. In particular, we present our method for verifying the correctness of the workflow output given the workflow and its metadata. General notations that are used in the proposed framework are presented in Table~\ref{table:notation} in Appendix~\ref{app:notation}.

\subsection{GWAS Statistics}\label{sec:gwas}

GWAS is a method used to identify genetic variations that are associated with a particular phenotype (usually a disease). In a typical GWAS, the researcher quantifies the associations between a disease (or phenotype) and a SNP using the odds ratio and the corresponding $p$-value. For a 2x2 contingency table, the odds ratio (OR) is computed as $OR = \frac{(S_{1,2})/(C_{1,2})}{S_0/C_0} = \frac{C_0 (S_{1,2})}{S_0 (C_{1,2})}$ (using the values in Table~\ref{table:gwas}). The $95\%$ confidence interval is from $\exp(ln(OR)-1.96\times SE\{ln(OR)\})$ to $\exp(ln(OR)+1.96\times SE\{ln(OR)\})$, where the standard error (SE) of the log odds ratio is $SE\{ln(OR)\}=\sqrt{\frac{1}{S_{1,2}} + \frac{1}{S_0} + \frac{1}{C_{1,2}} + \frac{1}{C_0}}$. The $p$-value is computed as in~\cite{sheskin2004inferential}. First, the standard normal deviation ($z$-value) is computed as $ln(OR)/ \\ SE\{ln(OR)\}$, and then the $p$-value is computed as the area of the normal distribution that falls outside $\pm z$. The $p$-value shows if the association between a SNP and a phenotype is statistically significant or not. SNPs whose $p$-values are low enough (smaller than a threshold) are considered significant. Finally, the minor allele frequency (MAF) of the SNPs in the case group is computed as $MAF = \frac{S_1+2\times S_2}{2\times S}$ (using the values in Table~\ref{table:gwas}). For the SNPs that have a $p$-value smaller than a threshold, the researcher publicly releases their aggregate statistics ($p$-value, OR, and MAF).

\subsection{GWAS and Generation of Metadata}\label{sec:gwas_usecase}

Let $D$ represent the dataset owned by the researcher. We denote the total number of individuals in the dataset (that are also involved in GWAS) as $n$, and the total number of SNPs as $m$. In the rest of this paper, we assume that the number of case and control users are equal ($n/2$ cases and $n/2$ controls). We let $t$ denote the phenotype being studied. For each SNP $j$, the researcher computes its association with phenotype $t$ in terms of the odds ratio ($o_j^t$) and $p$-value ($p_j^t$). The researcher also computes the minor allele frequency ($a_j^t$) of each SNP $j$ in the case group. As a result of GWAS, the researcher provides to the verifier the statistics of the $l$ most associated SNPs together with the SNP IDs as $R^t=\{R_1^t, R_2^t, \ldots, R_l^t\}$, where $R_j^t= \{o_j^t, p_j^t, a_j^t\}$. Note that sharing the IDs and (non-noisy) statistics of the most associated top-$l$ SNPs as a result of GWAS is required and allowed by many institutions, (e.g., NIH~\cite{nih}). In Section~\ref{sec:dp}, we also consider the scenario where statistics are shared under DP. The input of the workflow consists of the case/control groups, whereas the workflow output consists of the summary statistics ($p$-values, odds ratios, and MAFs) for the strongly associated SNPs with the studied phenotype.

By using the workflow along with the metadata received from the researcher, the verifier can check the correctness of the provided statistic for each SNP. Thus, as metadata, the researcher provides (i) the phenotype that is being studied, (ii) the population of the individuals in the GWAS study, (iii) the number of individuals in the case/control groups ($n$), and (iv) the number of SNPs in the study ($m$). As part of the metadata, the researcher also provides raw data (i.e., part of the original dataset) for the $k$ most significant SNPs after adding noise to it using the randomized response mechanism to achieve LDP (with parameter $\epsilon$), as discussed in Section~\ref{sec:ldp} along with the $\epsilon$ value. We denote this partial noisy dataset as $D_k^{\epsilon}$ and we assume that $k$ is greater than or equal to $l$. Briefly, for each SNP $j$ in $D_k^{\epsilon}$, its original value (e.g., 0) is kept intact with probability $\textbf{p}$ and its value is flipped with probability $\textbf{q}$ (e.g., flipped to 1 or 2 with equal probability), where $\textbf{p}=\frac{e^{\epsilon}}{2+e^{\epsilon}}$ and $\textbf{q}=\frac{1}{2+e^{\epsilon}}$. We opt for randomized response since the total number of potential values for each SNP is 3 and this mechanism provides the best utility for such few number of states~\cite{wang2017locally}. Moreover, the randomized response mechanism uses the same set of inputs and outputs without an encoding, which allows the data collector to use perturbed data directly. We analyze the privacy guarantees of the proposed framework in Section~\ref{sec:mem_inf}. The researcher does not provide the noisy version of the entire dataset since we observe that membership inference attacks become more powerful as more SNPs (even after adding noise) are published, as shown in Section~\ref{sec:eval_mem_inf}. Furthermore, in Appendix~\ref{sec:dataset_partioning}, we discuss creating the partial dataset ($D_k^{\epsilon}$) via sampling to further reduce the privacy risk due to the partial noisy dataset. 

\subsection{Verification of the Workflow Output}\label{sec:verification}

Here, we introduce our methodology for verifying the correctness of the workflow output. The proposed framework consists of two main parts: (i) selection of cut-off points (i.e., threshold values to distinguish between correct and incorrect statistics) for each reported statistic using a publicly available dataset (described in Section~\ref{sec:cut_off}); and (ii) determining whether each provided statistic is computed correctly or not (described here). 

First, by using the aggregation technique in~\cite{wang2017locally}, the verifier estimates the actual occurrences of $0$, $1$, and $2$ for each SNP based on the received noisy dataset $D_k^{\epsilon}$ (as described in Section~\ref{sec:ldp}). Then, it performs GWAS on the estimated counts and calculates the statistics ($p$-value, odds ratio, and MAF) for the $k$ SNPs in $D_k^{\epsilon}$, denoted as $Q^t=\{Q_1^t, Q_2^t, \ldots, Q_k^t\}$, where $Q_j^t= \{\hat{o}_j^t, \hat{p}_j^t, \hat{a}_j^t\}$. Note that the researcher does not provide $Q_k^t$ instead of $D_k^{\epsilon}$ because we assume that the researcher might make computational errors (and such errors will likely be repeated if the researcher also computes $Q_k^t$ since they will be following the same methodology). Next, the verifier computes the deviation (or distance) of the computed statistics (in $Q^t$) from the $l$ statistics provided by the researcher in $R^t$. To compute the deviation, the verifier uses the relative error (RE) metric computed between the statistics in $R^t$ and in $Q^t$. For the $p$-value, the verifier computes its deviation, denoted as $RE_p=\{RE_{p_1}, RE_{p_2}, \ldots, RE_{p_l}\}$, as the fraction of the logarithm of $p$-values that is lost due to the randomized response mechanism.
Formally, $RE_{p_j}$ is computed as:  
$$RE_{p_j}=\frac{|-ln(p_j^t) - (-ln(\hat{p}_j^t))|}{-ln(p_j^t)},$$  
where $j \in \{1, \ldots, l\}$. 
For the odds ratio, $RE_{o_j}$ is computed as $RE_{o_j}=\frac{|o_j^t - \hat{o}_j^t|}{o_j^t}.$ The deviation of the minor allele frequency ($RE_{a}$) is also computed similarly.

\begin{figure*}[ht!]
    \centering
    \includegraphics[scale=0.4]{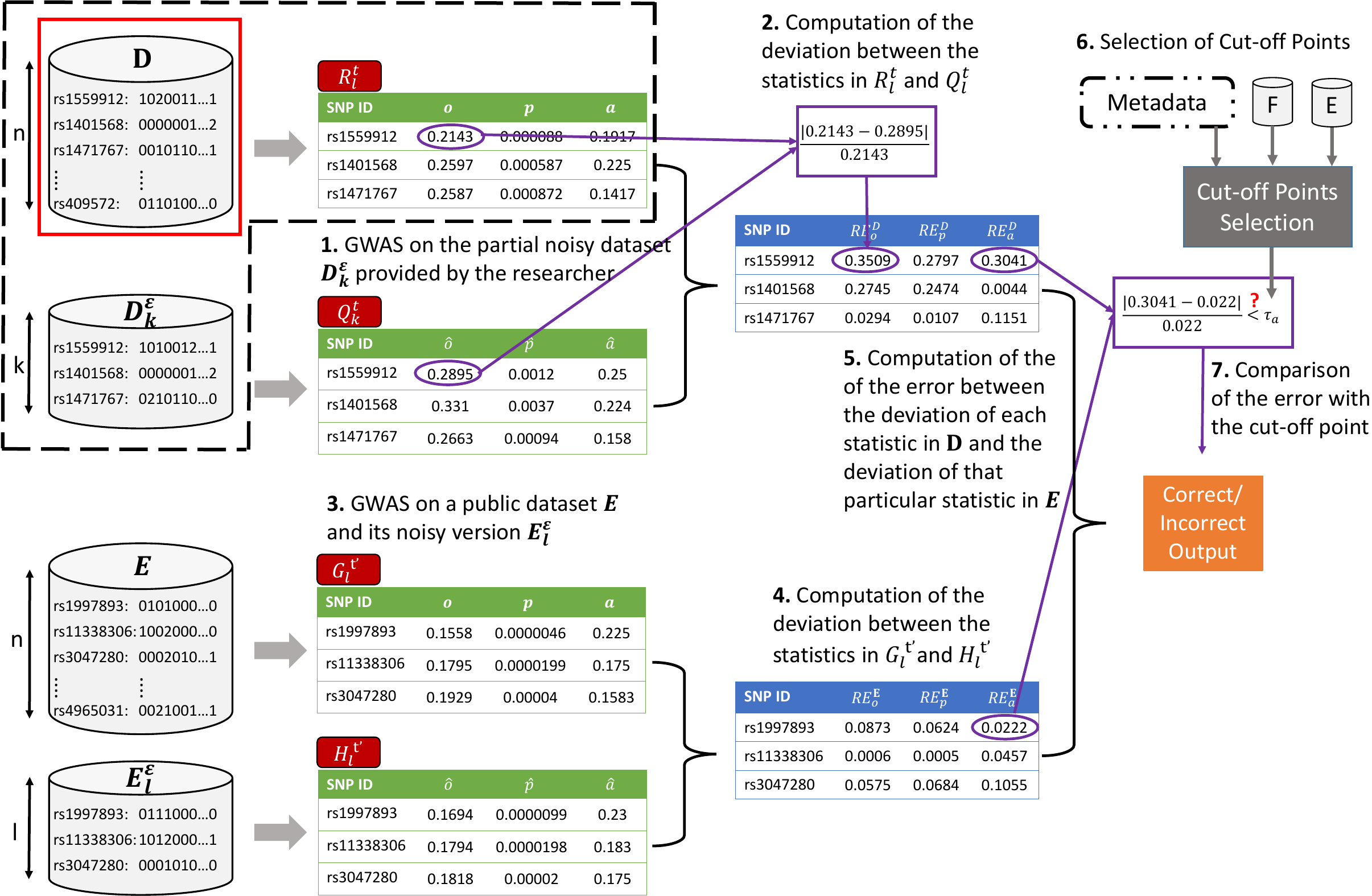}
    \caption{A toy example showing how the verifier checks whether the released statistics are correct or not. Some of the steps can be done in parallel, numbering is for clarity. $D$ denotes the researcher's (private) dataset, $E$ is the public dataset, and $F$ is another labelled public dataset. $D_k^{\epsilon}$ denotes the partial noisy dataset provided as part of the metadata and $E_l^{\epsilon}$ is the partial noisy public dataset. $R_l^t$ denotes the statistics provided by the researcher, $Q_k^t$ is the statistics obtained as a result of GWAS on $D_k^{\epsilon}$, $G_l^{t'}$ is the statistics obtained as a result of GWAS on $E$, and $H_l^{t'}$ is the statistics obtained as a result of GWAS on $E_l^{\epsilon}$. As the released GWAS statistics, $o$ denotes odds ratio, $p$ is the $p$-value, and $a$ is the minor allele frequency. Furthermore, $RE_{o}^D$ denotes the deviation of odds ratio in $D$, $RE_{p}^D$ is the deviation of $p$-value in $D$, and $RE_{a}^D$ is the deviation of minor allele frequency in $D$. Finally, $\tau_a$ denotes the cut-off point for minor allele frequency. $R_l^t$ and $D_k^{\epsilon}$ are provided to the verifier by the researcher.}
    \label{fig:example}
    \vspace{-10pt}
\end{figure*}
Here, our main assumption (as we validate in Section~\ref{sec:results}) is that independent of the phenotype being studied, the distance obtained between the statistics computed over $D_k^{\epsilon}$ and $D$ follows a similar trend for different datasets for a given $\epsilon$ value. To utilize this, the verifier computes the ``expected distance'' by using a publicly available genomic dataset $E$ and its noisy version $E^{\epsilon}$, in which each data point is obfuscated (by the verifier) using randomized response to achieve LDP (with the same privacy parameter that is used by the researcher, $\epsilon$). Using the previously introduced distance metrics (for researcher's statistics), the verifier computes the expected distance ($RE_{p}^E$ for $p$-value, $RE_{o}^E$ for odds ratio, and $RE_{a}^E$ for MAF) for the $l$ most associated SNPs from $E$ (denoted as $G^{t'}=\{G_1^{t'}, G_2^{t'}, \ldots, G_l^{t'}\}$) and $E^{\epsilon}$ (denoted as $H^{t'}=\{H_1^{t'}, H_2^{t'}, \ldots, H_l^{t'}\}$). The verifier does not need a labelled dataset for this. Instead, they can randomly label the dataset to compute the statistics. Here, the important point is that $E$ contains the same number of case and control users as in $D$.

After computing the expected distance of each statistic (i.e., $RE_{p}^E$, $RE_{o}^E$, and $RE_{a}^E$) on the public genomic dataset, the verifier computes the relative change, error, (e.g., $\Phi_{p_j}=\frac{|RE_{p_j}^D - RE_{p_j}^E|}{RE_{p_j}^E}$) between the deviation of each statistic in $D$ (e.g., $RE_{p_j}^D$) and the deviation of that particular statistic in $E$ (e.g., $RE_{p_j}^E$). If this change is smaller than a predefined cut-off point (threshold value), then the statistic provided by the researcher (e.g., $p_j$) is classified as being computed correctly; otherwise it is classified as incorrect. In Figure~\ref{fig:example}, we provide a toy example showing how the verification of received statistics is done. For each statistic ($p$-value, odds ratio, and MAF), the verifier can heuristically set a cut-off point ($\tau_o$, $\tau_p$, and $\tau_a$, respectively) depending on the error value (relative change from the expected distance of each statistic) they choose to tolerate. In the next section, we describe how to compute the cut-off points for each statistic by using a public dataset. 

\subsection{Selection of Cut-off Points}\label{sec:cut_off}

To compute the cut-off points, the verifier can use another labelled (publicly available) genomic dataset $F$, whose phenotype does not need to be the same as the one in $D$. The verifier uses $F$ to simulate the potential computational errors done by the researcher (with dataset $D$). In contrast to $D$, which is unknown to the verifier (except for its partial noisy dataset $D_k^{\epsilon}$), the verifier knows the ground-truth for each possible considered scenario in $F$ (i.e., whether the correct statistics are provided or not). Therefore, for each statistic, the verifier computes the probability distributions of the error between the statistic's deviation in $F$ (e.g., $RE_{p}^F$) and the statistic's deviation in $E$ (e.g., $RE_{p}^E$) when the correct and incorrect statistics are provided. For incorrect statistics, the verifier might consider different scenarios depending on how much the provided statistics' values deviate from the correct ones. Based on these distributions, the verifier identifies the points, at which both the false positive and false negative probabilities are minimized and selects these as the cut-off points for each statistic ($\tau_o$, $\tau_p$, and $\tau_a$, respectively). Here, a false positive is the outcome, for which an incorrect statistic is classified as correct, and a false negative is the outcome, for which a correct statistic is classified as incorrect. In order to avoid the dependence of the cut-off points on a single case-control setup ($F$), the verifier can partition $F$ into multiple splits (case-control setups), such that the SNPs in these splits do not intersect with each-other and use them to compute the cut-off points.

The performance of the proposed framework mainly depends on the number of statistics returned by the researcher ($l$) and the amount of noise added to the partial dataset ($\epsilon$). Thus, as we show in Section~\ref{sec:results}, the researcher can fine-tune these two parameters based on the verifier's confidence and the privacy level of the participants in researcher's dataset.

\section{Evaluation}\label{sec:results}

In this section, we evaluate the proposed framework by using real genomic data. We also study the impact of several factors to (i) the verifier's confidence for the verification of the GWAS output and (ii) the privacy of individuals in the researcher's dataset.

\subsection{Datasets and Evaluation Metrics}\label{sec:datasets}

We use two different genomic datasets for evaluation: (i) the OpenSNP dataset~\cite{opensnp}, which also includes phenotype information about the individuals, and (ii) the 1000 Genomes Phase 3 dataset~\cite{1000genomes}. We assume that the researcher's dataset $D$ is the OpenSNP dataset, while the 1000 Genomes dataset is a public dataset ($E$). From OpenSNP, we use the following phenotypes: (i) lactose intolerance ($D1$), (ii) hair color ($D2$), and (iii) handedness ($D3$). For each phenotype, we extract the genomic data of $120$ randomly selected individuals ($n=120$) where $60$ of them carry the trait (phenotype) and $60$ do not. From the OpenSNP dataset, we also extract genomic data of another $120$ individuals (that do not overlap with $D1$, $D2$, or $D3$) and create a case-control setup ($F$, which is used to compute the cut-off points, as discussed in Section~\ref{sec:cut_off}) consisting of $44000$ SNPs. To construct $E$ from the 1000 Genomes dataset, we extract the SNPs from chromosome 22 of $120$ randomly selected individuals from the European population (CEU). We randomly label the 1000 Genomes dataset and create a case-control setup consisting of $41000$ SNPs.  

To simulate a researcher that makes unintentional computational errors, first, we conduct GWAS on $D$ ($D1$, $D2$, and $D3$ in our experiments). Then, we sort the SNPs based on their increasing $p$-values. To simulate the overselling scenario, we assume that the $v$-th SNP in the sorted list is the one with the strongest association (with the lowest $p$-value) in the researcher's dataset. We use the $p$-values of SNPs from $1$ to $(v-1)$ in the sorted list as the incorrect values reported by the researcher. Thus, the $p$-value of the $(v-1)$-th SNP is the closest to the correct values, and the $p$-value of the first SNP is the one that deviates the most. For the statistics reported by the researcher, we pick $l$ consecutive $p$-values that have different deviations from the correct ones, and hence for each scenario, there are $l$ statistics that need to be classified (as correct or incorrect). We follow the same strategy for the other two statistics.

To quantify the success of the proposed verification scheme (i.e., verifier's confidence), for each statistic, we use true positive rate ($TPR = \frac{TP} {TP + FN}$) and true negative rate ($TNR = \frac{TN} {TN + FP}$). We consider a true positive ($TP$) as the outcome, in which the provided statistic is correctly classified as being correct; a false positive ($FP$) as the outcome, in which an incorrect statistic is classified as correct; a false negative ($FN$) as the outcome, in which a correct statistic is classified as incorrect; and a true negative ($TN$) as the outcome, in which an incorrect statistic is classified as being incorrect. Furthermore, to quantify the impact of the computational errors done by the researcher, we evaluate the utility loss of each statistic as a result of the provided incorrect values. We compute this as the distance between the statistics provided by the researcher ($R^t$) and the ones that should have been returned as part of the research (the correct statistics for the same SNP). For instance, for $p$-value, we compute its utility loss as $$U_p = \frac{1}{l} \sum_{j=1}^l \left( \frac{|p_j - y_j|}{Z} \right ),$$ where $p_j$ is the $p$-value of SNP $j$ released by the researcher, $y_j$ is the $p$-value of SNP $j$ when the computation is done correctly, and $Z$ is a normalization constant representing the maximum value that the $p$-value can take. The utility loss of odds ratio ($U_o$) and minor allele frequency ($U_a$) are also computed similarly.

\subsection{Verification Confidence}\label{sec:eval_verification}

The verification confidence provided by the proposed framework can be studied theoretically and empirically.

\subsubsection{Experimental Results}\label{sec:experimental_results}

To compute the cut-off points, we first randomly split $F$ into $5$ disjoint case-control studies. There is no overlap between the SNP IDs of any of these case-control setups. By following the technique described in Section~\ref{sec:cut_off}, we use these case-control setups and the one created from the 1000 Genomes dataset ($E$) to compute the cut-off points for each statistic ($\tau_o$, $\tau_p$, and $\tau_a$, respectively). In the rest of this section, we assume that the verifier always uses these case-control setups to determine the cut-off points for each $l$ and $\epsilon$ value. 

\begin{figure*}[ht!]
    \centering
    \begin{subfigure}[Lactose Intolerance ($D1$)]{\includegraphics[scale=0.38]{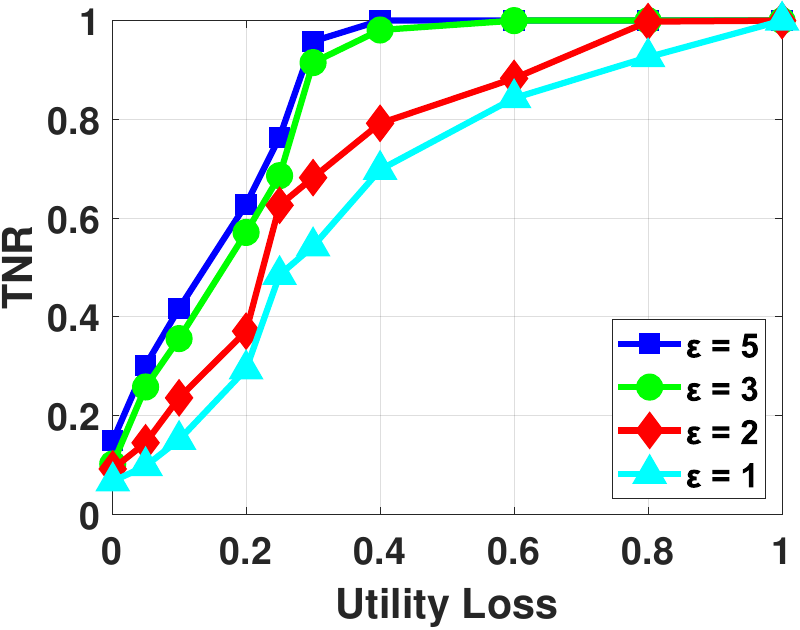}}
    \end{subfigure}\hfill
    \begin{subfigure}[Hair Color ($D2$)]{\includegraphics[scale=0.38]{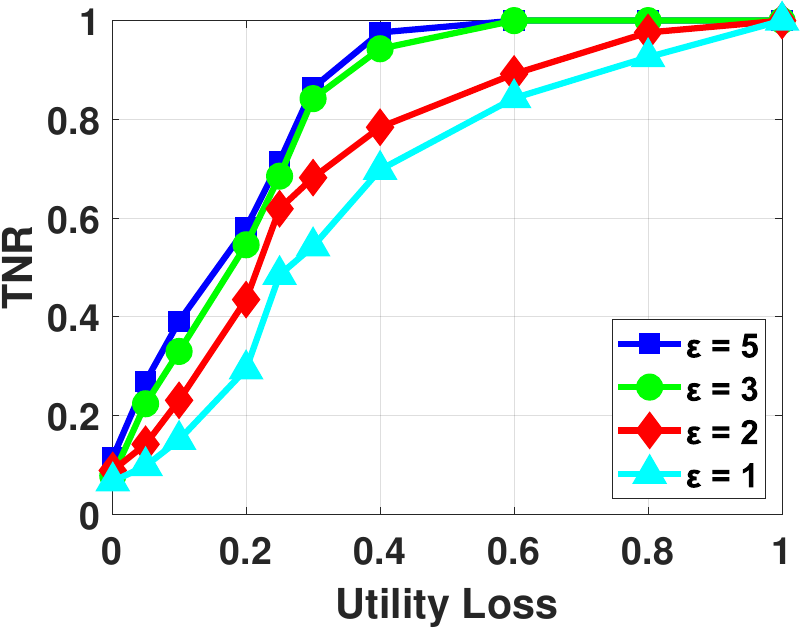}}
    \end{subfigure}\hfill
    \begin{subfigure}[Handedness ($D3$)]{\includegraphics[scale=0.38]{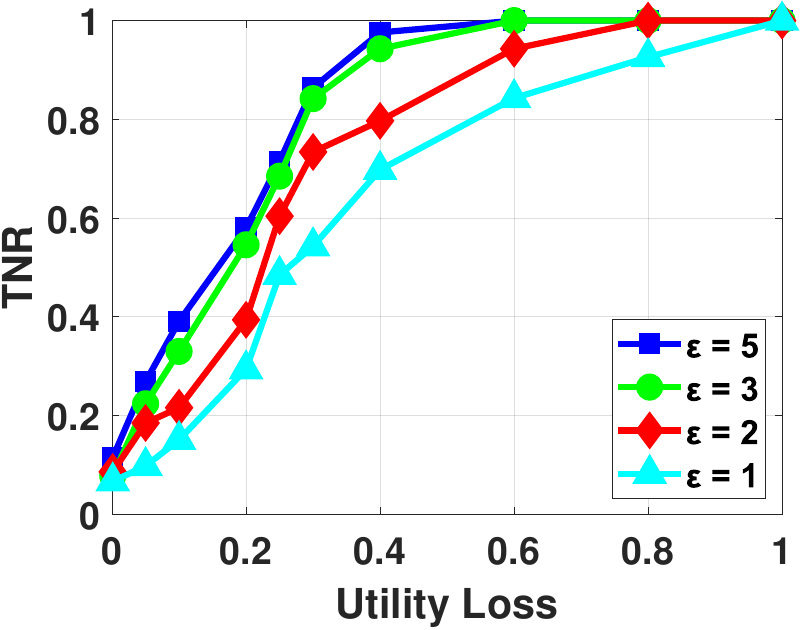}}
    \end{subfigure}\hfill
    \caption{Variation of TNR in the verification of $p$-value with respect to its utility loss for different $\epsilon$ values, when the number of released statistics is $100$ in $D1$, $D2$, and $D3$.}
    \label{fig:tnr_vs_eps_p_value}
    \vspace{-10pt}
\end{figure*}

We evaluate the proposed framework using the case-control setups of the OpenSNP dataset ($\mathrm{D1}$, $\mathrm{D2}$, and $\mathrm{D3}$). We assume that the researcher provides the top $l$ most significant SNPs and their corresponding statistics and that $k$, i.e., the size of the provided partial dataset, is equal to $l$. Note that the researcher provides raw data ($D_k^{\epsilon}$) only for the independent SNPs (as will be discussed in Section~\ref{sec:mem_inf}). Here, we focus on the overselling scenario, where the researcher (unintentionally) might provide stronger associations than the actual ones. Since there is randomness due to the noise addition via randomized response, we repeat each experiment $5$ times and report the average of the results. 

\textbf{Effect of $\epsilon$ value.} 
First, we study the impact of $\epsilon$ value, i.e., the amount of noise added to the original dataset (via randomized response) by the researcher before providing it as part of the metadata ($D_k^{\epsilon}$), to the TPR and TNR values. We fix $l$ (number of returned SNPs) to $100$ and consider $\epsilon$ values in $\{1, 2, 3, 5\}$. In Table~\ref{table:tpr_vs_eps}, we show the TPR values obtained using the proposed framework for each statistic and for each dataset ($D1$, $D2$, and $D3$) with varying values of $\epsilon$. We observe a TPR of $0.73$ (for $D1$), $0.69$ (for $D2$), and $0.67$ (for $D3$) for $p$-value when $\epsilon=3$ and when all returned statistics are correct (as shown in Table~\ref{table:tpr_vs_eps}). For $D1$, this means that the verifier can correctly classify $73$ out of $100$ statistics provided by the researcher. 
\begin{table}[ht!]
    \centering
    \vspace{-8pt}
    \caption{TPR for verifying the correctness of each statistic for varying values of $\epsilon$ when the number of released statistics is $100$ in $D1$, $D2$, and $D3$.}
    \vspace{-5pt}
    \begin{tabular}{|c|c|c|c|c|c|}
        \hline
        Dataset & statistic & $\epsilon=1$ & $\epsilon=2$ & $\epsilon=3$ & $\epsilon=5$ \\ \hline
        \multirow{3}{*}{$D1$} & $o$ & $0.45$ & $0.56$ & $0.62$ & $0.78$ \\ \cline{2-6}
        & $p$ & $0.47$ & $0.65$ & $0.73$ & $0.9$ \\ \cline{2-6}
        & $a$ & $0.44$ & $0.62$ & $0.7$ & $0.86$ \\ \hline
        \multirow{3}{*}{$D2$} & $o$ & $0.49$ & $0.55$ & $0.6$ & $0.78$ \\ \cline{2-6}
        & $p$ & $0.39$ & $0.59$ & $0.69$ & $0.87$ \\ \cline{2-6}
        & $a$ & $0.38$ & $0.6$ & $0.66$ & $0.84$ \\ \hline
        \multirow{3}{*}{$D3$} & $o$ & $0.44$ & $0.53$ & $0.59$ & $0.77$ \\ \cline{2-6}
        & $p$ & $0.43$ & $0.61$ & $0.67$ & $0.86$ \\ \cline{2-6}
        & $a$ & $0.37$ & $0.57$ & $0.67$ & $0.85$ \\ \hline
    \end{tabular}
	\label{table:tpr_vs_eps}
	\vspace{-5pt}
\end{table}

In the following, we assume that all returned statistics are incorrectly reported by the researcher to compute the TNR of the proposed scheme. We show the variation of the TNR values with respect to the utility loss for each statistic and each dataset (phenotype) in Figure~\ref{fig:tnr_vs_eps_p_value}, and Figures~\ref{fig:tnr_vs_eps_or}~-~\ref{fig:tnr_vs_eps_maf} in Appendix~\ref{app:varying_eps}. The proposed framework achieves a TNR of $0.57$ and $0.98$ (for $D1$) for $p$-value when $\epsilon = 3$ and when the utility loss is $0.2$ and $0.4$, respectively. As discussed, higher utility loss implies that the researcher deviates more from the correct values of the statistics. To understand more clearly what these results actually mean, consider an illustrative scenario, in which the researcher has obtained (as a result of GWAS) a $p$-value that is in the range $[.09-.1]$ and tries to oversell it by incorrectly reporting the $p$-value in the range $[.035-.045]$ (i.e., reporting a weak association as strong). Note that an association with a $p$-value below $.05$ is typically accepted as a strong one. Thus, the $p$-value threshold is $.05$. However, higher $p$-value thresholds may also result in false positives in GWAS results, and hence later in this section, we also study how the performance of the proposed framework changes with varying values of the $p$-value threshold. In this case, the utility loss is at least $0.35$ and the verifier can successfully detect this incorrect result with a probability close to $1$. For the other scenario, when a correct $p$-value is smaller than $.01$ and the researcher tries to report it in the range $[.05-.06]$ (i.e., reporting a strong association as weak), the utility loss is at least $0.32$ and again, the verifier can successfully detect that the provided $p$-value is incorrect with probability close to $1$. We also observe that as the deviation of the returned statistics ($p$-value, odds ratio, and minor allele frequency) from the correct ones (utility loss) increases, TNR also increases. For instance, for $\epsilon=3$, a verifier can determine with high confidence (a TNR of at least $0.8$) if the researcher has released incorrect $p$-values when the utility loss is at least $0.28$. Whereas, the verifier obtains a low TNR (smaller than $0.4$) when the released $p$-values are close to the correct ones (when the utility loss is less than $0.15$). In such cases, even though the verifier's success in classifying the returned $p$-values as incorrect is low, the returned $p$-values still preserve a high utility. Therefore, the proposed framework successfully (and with high confidence) detects when a weakly associated SNP is reported to have a low $p$-value and vice versa. We obtained similar results for the other two datasets $D2$ and $D3$. We also observe that as $\epsilon$ value decreases (i.e., the amount of noise added to the researcher's dataset increases), TNR decreases. As discussed in Section~\ref{sec:eval_mem_inf}, a researcher can set $\epsilon$ based on the privacy risk of the research participants due to the partial noisy dataset.

The results we showed so far do not consider the $p$-values of the returned statistics (they only consider the classification accuracy of the verifier for the returned statistics). On the other hand, in practice, the verifier is only interested in strongly associated SNPs with $p$-values smaller than $.05$. 
\begin{figure}[ht!]
    \centering
    \begin{subfigure}[Correct Statistics]{\includegraphics[scale=0.29]{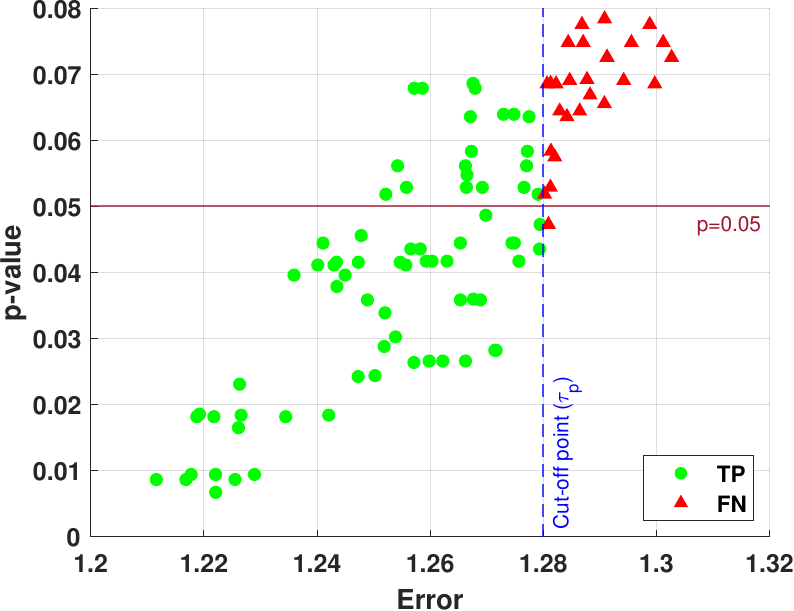}}
    \end{subfigure}\hfill
    \begin{subfigure}[Incorrect Statistics]{\includegraphics[scale=0.29]{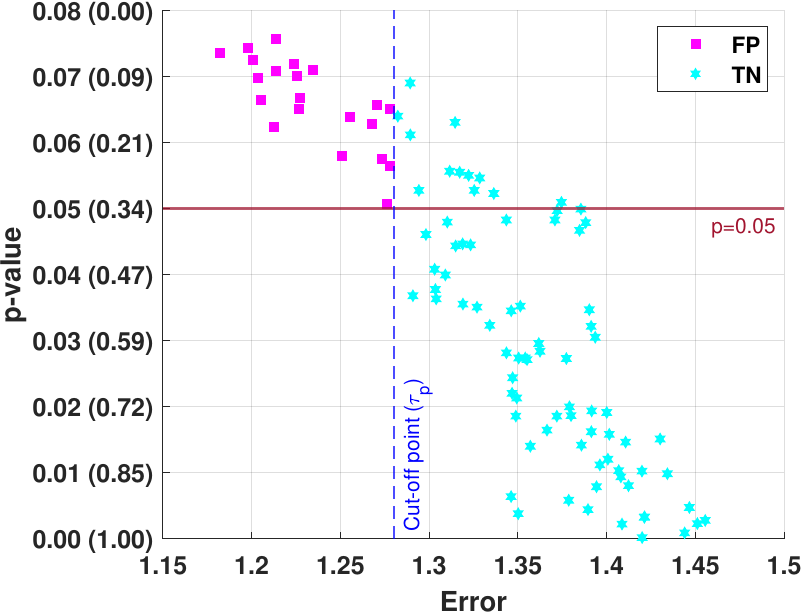}}
    \end{subfigure}\hfill
    \caption{Error of the $p$-values ($\Phi_p$) with respect to their reported $p$-values for both correct and incorrect statistics for $\epsilon=3$ in $D1$. For the incorrect statistics, the values in parenthesis (on the y-axis) show the utility loss for the corresponding reported $p$-value when its correct $p$-value is $0.08$.}
    \label{fig:error_vs_p_value}
    \vspace{-10pt}
\end{figure} 
In Figure~\ref{fig:error_vs_p_value}(a), we show the error ($\Phi_p$) obtained by the verifier for $100$ correctly computed statistics ($p$-values) by the researcher as a result of GWAS with respect to $\tau_p$ and their $p$-values for $\epsilon=3$. We observe that the verifier achieves a TPR of almost 1 for the strong associations (i.e., the ones with $p$-values smaller than $.05$). Also, in Figure~\ref{fig:error_vs_p_value}(b), we show the error of $100$ incorrect statistics and their reported $p$-values, whose correct $p$-value is around $.08$. We observe that the verifier obtains a TNR of almost 1 when the researcher (erroneously) oversells a weak association as a strong one. In studies with multiple hypothesis, for higher $p$-value thresholds (e.g., $.05$), it is likely to have false positive GWAS findings by identifying incorrect SNPs as the associated ones. A common way to reduce the probability of such false positives and counteract the problem of multiple comparisons is to use Bonferroni correction, which is also commonly used in GWAS settings~\cite{kuo2017multiple}. Therefore, we evaluate the performance of the proposed framework with varying $p$-value thresholds. We obtain similar results when we reduce the $p$-value threshold for strong associations from $.05$ to $.01$ or $5.5 \times 10^{-6}$ (computed by using Bonferroni correction). These results show that a verifier can use the proposed scheme with high confidence, especially for the SNPs that are highly associated with the studied phenotype for varying values of the p-value threshold \textbf{that are below $.05$}. Therefore, even if a researcher uses multi-test adjustment methods (such as Bonferroni correction) to minimize the false positives in its findings, the performance of the proposed scheme remains intact.

\textbf{Effect of the number of returned statistics.} Next, we fix the $\epsilon$ value to $5$ and vary the number of returned statistics by the researcher ($l$) for $l \in \{10, 50, 100, 200\}$. We observe that the number of statistics returned by the researcher ($l$) does not have a significant effect in TPR and TNR. Due to space constraints, we show the results of this experiment in Table~\ref{table:tpr_vs_l} and Figure~\ref{fig:tnr_vs_l_p_value}, in Appendix~\ref{app:varying_l}. The researcher can decide on the number of returned SNPs ($l$) by considering the vulnerability of the research participants due to the membership inference attacks (as discussed in Section~\ref{sec:eval_mem_inf}).

\begin{figure}[ht!]
    \centering
    \includegraphics[scale=0.38]{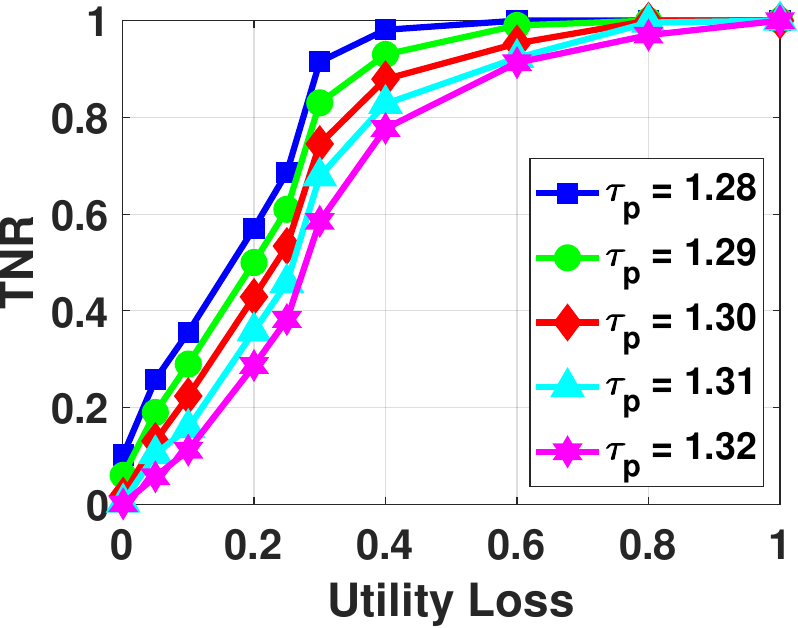}
    \caption{Variation of TNR in the verification of $p$-value with respect to its utility loss for different cut-off points ($\tau_p$ values), when $\epsilon=3$ and $l=100$ in $D1$.}
    \label{fig:cut-off_points}
    \vspace{-10pt}
\end{figure}

\begin{wraptable}{R}{0.2\textwidth}
    \centering
    \vspace{-10pt}
     \caption{TPR for verifying the correctness of $p$-value for different cut-off points ($\tau_p$ values), when $\epsilon=3$ and $l=100$ in $D1$.}
     \vspace{-5pt}
    \begin{tabular}{|c|c|}
        \hline
        Cut-off Point & TPR \\ \hline
        $1.28$ & $0.73$\\ \hline
        $1.29$ & $0.78$\\ \hline
        $1.30$ & $0.83$\\ \hline
        $1.31$ & $0.89$\\ \hline
        $1.32$ & $0.95$\\ \hline
    \end{tabular}
	\label{table:tpr_vs_cut-off}
	\vspace{-10pt}
\end{wraptable}
\textbf{The effect of cut-off points.} Another parameter that affects the performance of the proposed framework (especially to TPR) is the cut-off points selection. Here, we explore the effect of the cut-off points to the TPR and TNR values. For our evaluation, we use $D1$ and set $l=100$ and $\epsilon=3$. The cut-off point, $\tau_p=1.28$, is computed by using datasets $E$ and $F$. Table~\ref{table:tpr_vs_cut-off} shows the TPR values achieved by the proposed framework when verifying the correctness of $p$-value using different cut-off points and Figure~\ref{fig:cut-off_points} shows the variation of TNR with respect to the utility loss for different cut-off points. We obtain similar results (trends) for the other $\epsilon$ values and statistics. As the cut-off point ($\tau_p$) increases, TPR increases while TNR decreases. However, we observe that the decrease in TNR is not significant considering that utility loss is small. Therefore, a verifier can use a higher cut-off point to have a high TPR performance without compromising much in terms of TNR and utility.

\textbf{Considering error in metadata.} Apart from unintentional errors during the computation of the statistics, the researcher may also do errors during the generation of the partial noisy dataset $D_k^{\epsilon}$ (part of metadata, used for verification). To simulate this behavior, we consider a scenario where the researcher, instead of using the $\epsilon$ value they intend to use to achieve LDP ($\epsilon_x$), they use a smaller or larger value ($\epsilon_y$) to compute the partial noisy dataset $D_k^{\epsilon_y}$ and provide $\epsilon_x$ as part of the metadata (so that the verifier conducts the verification based on $\epsilon_x$). In other words, the researcher adds either more or less noise than they intend to add to generate the partial noisy dataset. For evaluation, we consider a researcher that does errors during both the computation of the statistics and the generation of the partial noisy dataset. Overall, we observe that proposed framework detects with high confidence whether the provided statistics are correct even when there exist (unintentional) errors in the metadata in addition to the errors in the calculation of the GWAS statistics. We observe that the TPR and TNR values are slightly lower than the ones achieved when the researcher computes $D_k^{\epsilon}$ correctly (when $\epsilon_y=\epsilon_x$ for different values of $\epsilon$).

\subsubsection{Theoretical Analysis}\label{sec:theoretical_analysis}

Here, we analytically examine the performance of the proposed verification framework. In particular, we explore the verification confidence and the robustness of the proposed scheme for the selection of the public datasets (e.g., $E$). In the following, we conduct our analysis considering a single SNP; the analysis can easily be generalized for multiple SNPs similarly. Let $y_j$ be the correct $p$-value of a SNP $j$ in dataset $D$, and $S_0$, $S_{1,2}$, $C_0$, $C_{1,2}$ be its corresponding counts values (as in Table~\ref{table:gwas}). Assume that the researcher (due to miscalculations) reports the $p$-value of SNP $j$ as $p_j = y_j + \eta$, where $\eta$ is the distance between the correct $p$-value and the reported one. Given the $\epsilon$ value (that will be known by the verifier), to achieve LDP via randomized response, we compute probabilities $\textbf{p}$ and $\textbf{q}$ (as described in Section~\ref{sec:ldp}). Thus, we compute the (expected) noisy counts of SNP $j$ as follows: $S_0^{\epsilon} = \textbf{p} S_0 + \textbf{q} S_{1,2}$, $S_{1,2}^{\epsilon} = \textbf{p} S_{1,2} + 2 \textbf{q} S_0$,  $C_0^{\epsilon} = \textbf{p} C_0 + \textbf{q} C_{1,2}$, and $C_{1,2}^{\epsilon} = \textbf{p} C_{1,2} + 2 \textbf{q} C_0$. Then, we compute $\hat{p_j}$ ($p$-value that is computed by the verifier using the received metadata) from the noisy counts, as described in Section~\ref{sec:gwas}. Let $\lambda_E$ denote the average distance between the $p$-values of the top $l$ SNPs in datasets $D$ and $E$ (i.e., $\lambda_E$ represents the distance between the research dataset $D$ and the public dataset $E$). Since in this analysis, we consider only one SNP in both datasets $D$ and $E$, we set $p_i^E = y_j + \lambda_E$, where $i$ is any of the top $l$ SNPs in $E$. We also compute $\hat{p_i}^E$ by using probabilities $\textbf{p}$ and $\textbf{q}$, as described before. For each dataset, we compute the relative error, $RE_{p_j}^D$ and $RE_{p_i}^E$ (as described in Section~\ref{sec:gwas_usecase}). Finally, we compute the error as $\Phi_{p} = \frac{|RE_{p_j}^D - RE_{p_i}^E|}{RE_{p_i}^E} = \frac{\frac{|-ln(p_j^D) - (-ln(\hat{p}_j^D))|}{-ln(p_j^D)}}{\frac{|-ln(p_i^E) - (-ln(\hat{p}_i^E))|}{-ln(p_i^E)}} - 1 = \frac{(|-ln(f(\psi) + \eta) + ln(f(\psi, \textbf{p}, \textbf{q}))|) ln(f(\psi) + \lambda_E)}{(|-ln(f(\psi) + \lambda_E) + ln(f(\psi, \textbf{p}, \textbf{q}) + \lambda_E)|) ln(f(\psi) + \eta)} - 1$, where $\psi = \{S_0, S_{1,2},\\ C_0, C_{1,2}\}$ and $f(\psi)$ is the function for computing the $p$-value of a SNP given its counts. Note that when the counts of a SNP are unknown (e.g., the noisy counts of SNP $j$), we include the parameters (e.g., probabilities $\textbf{p}$ and $\textbf{q}$) that we use to compute their (expected) values. So, the error (in this case $\Phi_p$) depends on the original counts ($\psi$), $\epsilon$ value, the distance between the reported value of the statistic and its original value ($\eta$), and the distance between $D$ and $E$ ($\lambda_E$).

\begin{figure}[ht!]
    \centering
    \begin{subfigure}[Incorrect Statistics]{\includegraphics[scale=0.29]{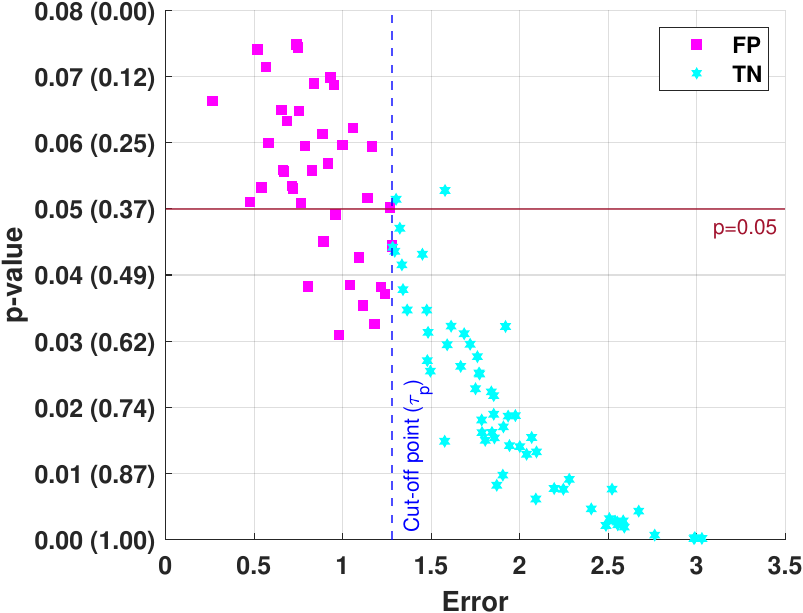}}
    \end{subfigure}\hfill
    \begin{subfigure}[Varying $E$]{\includegraphics[scale=0.29]{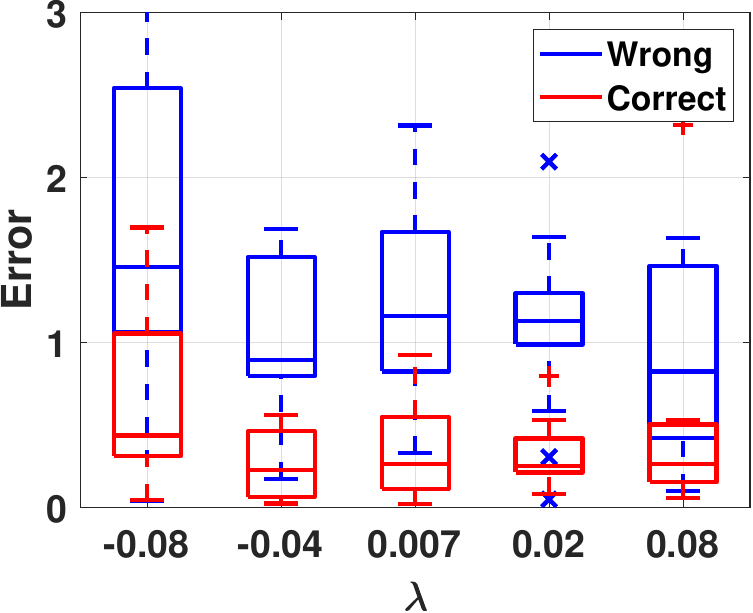}}
    \end{subfigure}\hfill
    \caption{Theoretical analysis. (a) Error of the $p$-values ($\Phi_p$) with respect to their reported $p$-values for the incorrect statistics for $\epsilon=3$ and $\lambda_E=0.01$. The values in parenthesis (on the y-axis) show the utility loss for the corresponding reported $p$-value when its correct $p$-value is $.08$. (b) Effect of the associations in $E$ to error ($\Phi_p$) for the verification of $p$-values for $\epsilon = 3$. Higher the gap between the error bars in correct and wrong statistics, better.}
    \label{fig:error_vs_p_value_theor}
    \vspace{-15pt}
\end{figure}

\textbf{Performance of the proposed framework. }First, we analyze the verification performance of the proposed framework. In Figure~\ref{fig:error_vs_p_value_theor}(a), we show the error of 100 incorrect $p$-values and their reported $p$-values whose correct $p$-value is around $0.08$ for $\epsilon=3$ and $\lambda_E=0.01$. Since the error values ($\Phi_p$) we obtained are similar to the ones we obtained in our empirical analysis (in Figure~\ref{fig:error_vs_p_value}), we use the same cut-off point as in our empirical analysis ($\tau_p=1.28$). Alternatively, we can also compute the cut-off points by modeling dataset $F$ via a similar analysis, as will be discussed later. We observe that via the theoretical analysis, one can achieve a TNR of almost 1 when the researcher oversells a weak association as a strong one, which is consistent with our empirical findings. We also did the same evaluation for the correctly reported $p$-values (all smaller than $.05$ to represent strong associations) and observe that the verifier achieves a TPR of 1. We obtain similar results for varying values of the $p$-value threshold for strong associations (e.g., using a $p$-value threshold of $.005$, which is computed after applying the Bonferroni correction).

\textbf{Effect of dataset $E$ selection. }Next, we explore the robustness of the proposed scheme with respect to the selection of public dataset $E$. That is, we explore the effect of the strength/weakness of the associations in $E$, the dataset used to compute the ``expected distance'' of each statistic to the performance of the verification framework. We focus on the overselling scenario, where a researcher reports a $p$-value of $.085$ as $.035$ for $\epsilon=3$. For $E$, we consider five scenarios with different $\lambda_E$ values. In Figure~\ref{fig:error_vs_p_value_theor} (b), we show the error ($\Phi_p$) achieved for each scenario. Overall, we observe that the performance of the proposed framework is robust for the selection of dataset $E$ within a large interval for $\lambda_E$. To show where some real-life datasets stand with respect to this interval, we analyze the datasets we used for $D$ and $E$ for our empirical evaluation (in Section~\ref{sec:experimental_results}). We observe that the 1000 Genomes dataset, which we used as dataset $E$ in Section~\ref{sec:experimental_results}, is within this interval with respect to dataset $D$ ($\lambda_E=0.005$). We simulate the above scenario by selecting SNPs from 1000 Genomes dataset whose distance to $D1$ are at the same $\lambda_E$ values. Our results support the theoretical analysis. We also obtained similar results for different $p$-values and $\eta$ values.

\textbf{Selection of cut-off points.} Similarly, we also explore the effect of dataset $F$ (and hence, the selection of cut-off points) to the performance of the proposed framework. Let $\lambda_F$ be the average distance between the $p$-values of the top $l$ SNPs in datasets $F$ and $D$. We compute the cut-off point for $p$-value as follows: 
\begin{enumerate}
    \item Given the $p$-value of a SNP $j$ in dataset $D$ (researcher's dataset) and its corresponding counts, compute its correct value $y_k^F = y_j + \lambda_F$, where $k$ is any of the top $l$ SNPs in $F$.
    \item Compute the reported $p$-value as $p_k^F = y_k^F + \eta$, where $\eta$ is the distance between the reported $p$-value and the correct one.
    \item Compute $\hat{p_k}^F$ by using probabilities $\textbf{p}$ and $\textbf{q}$ (as described at the beginning of Section~\ref{sec:theoretical_analysis}).
    \item Compute $RE_p^F$ (as described in Section~\ref{sec:gwas_usecase}).
    \item Compute $\Phi_p^F = \frac{|RE_{p_k}^F - RE_{p_i}^E|}{RE_{p_i}^E} = \\ \frac{(|-ln(f(\psi) + \lambda_F + \eta) + ln(f(\psi, \textbf{p}, \textbf{q}) + \lambda_F)|) ln(f(\psi) + \lambda_E)}{(|-ln(f(\psi) + \lambda_E) + ln(f(\psi, \textbf{p}, \textbf{q}) + \lambda_E)|) ln(f(\psi) + \lambda_F + \eta)} - 1$.
    \item Repeat steps 1-5 for different $\eta$ values.
    \item Compute the probability distributions of the error $\Phi_p^F$ when the correct and incorrect $p$-values are given.
    \item Select as the cut-off point $\tau_p$ the value which minimizes the false positive and negative probabilities.
\end{enumerate} 
Similar to before, we observe that the performance of the proposed framework is robust for the selection of dataset $F$ within a large interval for $\lambda_F$. We do not show results for this analysis due to space constraints.

\subsection{Privacy Analysis}\label{sec:mem_inf}

As discussed, the researcher provides a part of the raw dataset after adding noise to it using the randomized response mechanism to achieve LDP. To preserve the privacy guarantees of LDP, within $D_k^{\epsilon}$, the researcher provides data only for the SNPs that are independent from each-other based on public knowledge (i.e., linkage disequilibrium, LD).\footnote{Linkage disequilibrium is the publicly known pairwise correlations between the SNPs.} Via LDP, we provide indistinguishability between the original and reported values of the SNPs in $D_k^{\epsilon}$. For each SNP in $D_k^{\epsilon}$, the probability of distinguishing between the correct and reported value is bounded by $e^{\epsilon}$. Thus, LDP provides formal privacy guarantees against the attribute inference attack, but it does not provide such guarantees against the membership inference attack.

In the following, we quantify the power of the membership inference attacks due to the shared metadata and compare this privacy risk with the membership inference risk due to the shared GWAS statistics ($R^t$). As discussed, our goal is to show that the risk due to the proposed scheme does not increase the overall privacy risk due to the sharing of summary statistics, which is acceptable by many institutions, such as the NIH~\cite{nih}. We assume that the verifier has access to the victim's SNP profile, which can be extracted from a blood sample. To quantify the membership inference risk due the released statistics, we use the likelihood-ratio test (LRT) in Sankararaman et al.~\cite{sankararaman2009genomic}. A (misbehaving) verifier might attempt to determine whether a target victim is in the case group by computing the distance between the genome of the target victim and the partial noisy genomes of the users in the case group (part of $D_k^{\epsilon}$). We refer to this attack as ``hamming distance''. In the following, we briefly describe LRT to quantify the membership inference risk due to shared GWAS statistics and the hamming distance attack to quantify the membership inference risk due to the shared partial noisy dataset.

\subsubsection{Likelihood-Ratio Test}\label{sec:llr}

Let $x_{i,j}$ denote the SNP $j$ of individual $i$, where $x_{i,j} = \{0, 1, 2\}$, $a_j$ denote the aggregate allele frequency of SNP $j$ in the case group, and $pop_j$ the aggregate allele frequency of SNP $j$ in the reference population. The aggregate allele frequencies of the case group are provided as a part of the GWAS output, while the population allele frequencies can be acquired from public data sources, such as the 1000 Genomes project. Sankararaman et al.~\cite{sankararaman2009genomic} have empirically shown that, in the genomic setting, LRT is more powerful than the attack proposed by Homer et al.~\cite{homer2008resolving}, especially when the false-positive rate is low. We assume that under the null hypothesis, a target $i$ is not a part of the case group and under the alternate hypothesis, target $i$ is a part of the case group. The overall log-likelihood ratio can be computed as $LRT = \sum_{j=1}^l x_{i,j} log\frac{a_j}{pop_j} + (1-x_{i,j}) log\frac{1-a_j}{1-pop_j}.$

\subsubsection{Hamming Distance}\label{sec:edit_distance}

Here, a misbehaving verifier wants to find out whether any individuals' genome in the partial noisy dataset ($D_k^{\epsilon}$) provided as a part of the metadata is a match to that of a target victim. To identify the match (or closeness), we propose to use the hamming distance between the genomes. Hamming distance shows the minimum number of positions at which the genome sequences are different. For instance, given $X=GCTTACGA$ and $Y=GTTGACGA$, the minimum number of substitutions required to convert $X$ to $Y$ is $2$. In the following, we discuss the power analysis (for membership inference attack) using hamming distance between genomes. Assume the number of SNPs in $D_k^{\epsilon}$ is $k$. First, we use $|A|$ individuals from a set $A$ that are not in the case group of dataset $D$. For each individual in $A$, we compute the hamming distance between the target $i$ and all individuals in the case group of $D_k^{\epsilon}$ (only for the $k$ SNPs) and identify the minimum hamming distance. Then we identify the ``hamming distance threshold'' $\gamma$ as the 5\% false positive rate (for which 95\% of individuals in $A$ are correctly identified as not in the case group of $D$). Next, we use $|B|$ individuals from a set $B$ that are in the case group of dataset $D$. For each individual in $B$, we compute the hamming distance between the target $i$ and all individuals in the case group of $D_k^{\epsilon}$ and identify the minimum hamming distance. Finally, we check what fraction of these $|B|$ individuals have minimum hamming distance that is lower than the threshold $\gamma$ (i.e., correctly identified as in $D$), which gives the power of a misbehaving verifier.
 
\subsubsection{Results}\label{sec:eval_mem_inf}

\begin{figure}[ht!]
    \centering
    \vspace{-5pt}
    \includegraphics[scale=0.38]{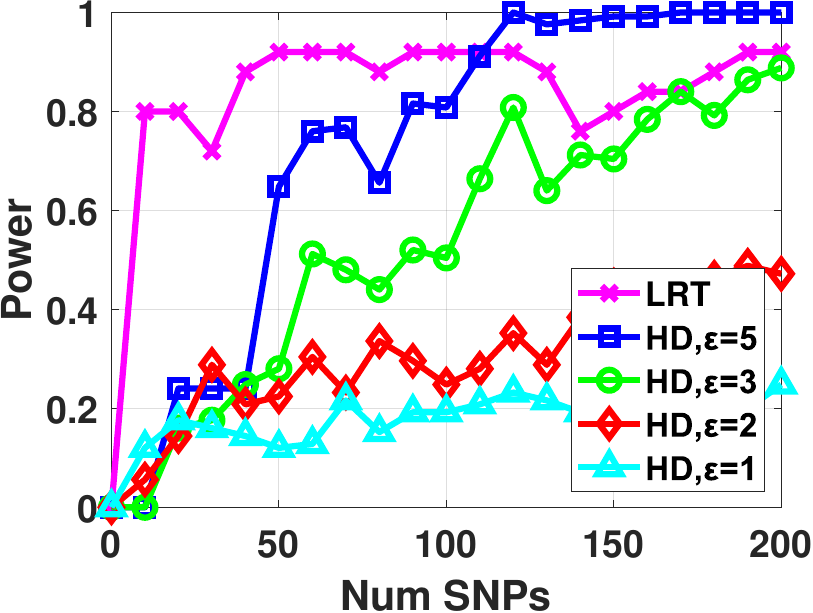}
    \caption{Power of the attacker for the membership inference attack for (i) different number of returned statistics ($l$) from $D1$ and (ii) varying number of SNPs ($k$) in the partial noisy dataset $D_k^{\epsilon}$ ($k=l$) for different $\epsilon$ values. HD stands for hamming distance.}
    \label{fig:mem_inf_or}
    \vspace{-15pt}
\end{figure}

We use $D1$ to evaluate the performance of membership inference attacks. We empirically build the null hypothesis using $|A|=25$ individuals that are not part of the case group. We reject the null hypothesis when LRT value is greater than a threshold value $\gamma$ (corresponding to $5\%$ false-positive rate) and when the minimum hamming distance is smaller than $\gamma$, respectively. For testing, we use $|B|=25$ randomly selected individuals from the case group of $D1$. Figure~\ref{fig:mem_inf_or} displays the power curve for LRT when the number of statistics ($l$) provided from $D1$ by the researcher varies. In the same figure, we also show the power curves for hamming distance for different $\epsilon$ values while varying the number of SNPs ($k$) in the partial noisy dataset $D_k^{\epsilon}$. For these experiments, we set $k=l$. As expected, as the number of provided statistics (MAF values of SNPs) or the number of SNPs in $D_k^{\epsilon}$ increases, the power of both attacks also increases. We observe that the power of the hamming distance attack on the partial noisy dataset is greater than the power of LRT on the released statistics only when $\epsilon=5$ and $D_k^{\epsilon}$ includes at least $60$ SNPs. For all other cases, the privacy risk due to the proposed framework is lower than the one due to the released GWAS statistics. Thus, we conclude that to preserve the privacy of dataset participants, the researcher can pick an $\epsilon$ value that is smaller than $5$. As we showed in Section~\ref{sec:eval_verification}, the verifier also achieves a high TPR and TNR for $\epsilon < 5$. 

\section{Discussion}\label{sec:discussion}

In this section, we discuss the complexity of the proposed scheme, its potential extensions, and limitations.

\subsection{Complexity}

The overall complexity of the proposed framework is dominated by the running time of GWAS, in other words, the computation of $Q_k^t$, $G_l^{t'}$, and $H_l^{t'}$ (steps 1 and 3 in Figure~\ref{fig:example}), and it can easily be parallelized (i.e., each SNP can be processed in parallel). We measured the running time in seconds with varying number of returned statistics ($l$). The proposed verification framework takes $0.69$ sec. to check the correctness of the statistics of $100$ SNPs, $1.14$ sec. for $200$ SNPs, $2.48$ sec. for $500$ SNPs, and $4.76$ sec. for $1000$ SNPs, respectively. Note that we do not process the SNPs in parallel. Also, a researcher typically returns only the statistics of the strong associations and that number is typically small. Thus, the proposed framework is practical and efficient.

\subsection{Considering Errors during Quality Control}\label{sec:qc}

As discussed, errors in the shared statistics may also occur during the quality control (QC) steps~\cite{turner2011quality,zuvich2011pitfalls}, which aim to eliminate noise and bias from the research dataset before GWAS. Similarly, the parameters used by the researcher for QC may not be enough to meet the quality standards of a client (another researcher). The proposed verification framework can also be used to determine if the research dataset is of high quality and identify the potential errors during the QC steps~\cite{turner2011quality} in a privacy-preserving way. Most of the sample QC steps, such as sample relatedness, population substructure and the marker QC steps (described in Appendix~\ref{app:qc_steps}) require statistical computations, and they can be (probabilistically) verified using the partial noisy dataset ($D_k^{\epsilon}$) provided as part of the metadata. For this, the verifier will first check if $D_k^{\epsilon}$ passes the quality control steps (the ones that can be statistically checked). If that is the case, then they will verify the correctness of the provided statistics following the steps in Figure~\ref{fig:example}. Otherwise, they will inform the researcher that the data is not of high quality. We will integrate the verification of QC steps into the proposed scheme in future work.

\subsection{Adding Noise to the GWAS Statistics}\label{sec:dp}

\begin{figure}[ht!]
    \centering
    \includegraphics[scale=0.38]{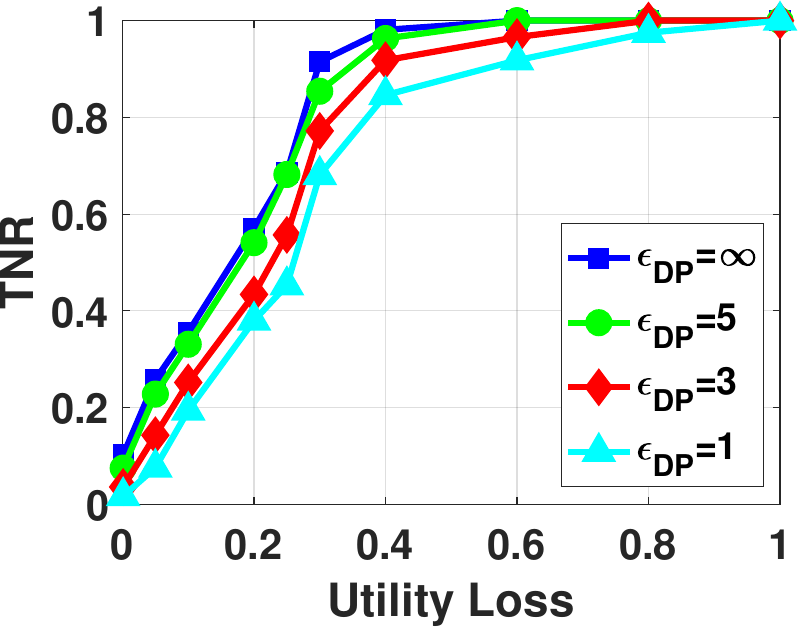}
    \caption{Variation of TNR in the verification of $p$-value with respect to its utility loss for different $\epsilon_{DP}$ values (noise added to the $p$-values by the researcher), when $\epsilon=3$ (noise added to generate $D_k^{\epsilon}$ under LDP) and $l=100$ in $D1$.}
    \label{fig:tnr_vs_eps_dp_p_value}
    \vspace{-15pt}
\end{figure}

In order to further protect the privacy of the research participants (especially if the research dataset includes a sensitive cohort), the researcher might add Laplacian noise to the statistics to achieve differential privacy - DP (with parameter $\epsilon_{DP}$) before publicly sharing them~\cite{johnson2013privacy,uhlerop2013privacy}. Here, we evaluate the performance of the proposed framework in such a setting. We fix $\epsilon$ (used to generate the partial noisy dataset $D_k^{\epsilon}$ under LDP) to $3$, $l$ to $100$, and consider $\epsilon_{DP}$ values from $\{1, 3, 5\}$. We observe that the proposed framework achieves a TPR of $0.4$ for $\epsilon_{DP}=1$, $0.57$ for $\epsilon_{DP}=3$, and $0.68$ for $\epsilon_{DP}=5$ when all the correct $p$-values are returned after noise addition. Figure~\ref{fig:tnr_vs_eps_dp_p_value} shows the variation of the TNR values with respect to the utility loss for $p$-value in $D1$. For all $\epsilon_{DP}$ values, TPR and TNR are lower than the ones obtained when no noise is added to the $p$-values (as shown in Table~\ref{table:tpr_vs_eps} and Figure~\ref{fig:tnr_vs_eps_p_value}(a)). We obtain similar results for the other two statistics: odds ratio and MAF. As expected, as $\epsilon_{DP}$ value decreases, TPR and TNR also decrease. At the same time, the utility of the provided statistics degrades especially for $\epsilon_{DP}=1$ (as also shown in~\cite{johnson2013privacy,uhlerop2013privacy}) rendering the reported statistics useless for other researchers.
As discussed, the verifier is interested in strongly associated SNPs (with $p$-values smaller than $.05$). Due to space constraints, we show the results for $\epsilon_{DP}=1$ and $\epsilon_{DP}=3$ in Appendix~\ref{app_dp}. We observe that the verifier can detect with high confidence whether the provided statistics are correct while ensuring that the privacy of the research participants is preserved. At the same time, the utility of the reported statistics is significantly reduced especially when $\epsilon_{DP}=1$.

\subsection{Limitations of the Proposed Framework}\label{sec:limitations}

As discussed, in the proposed framework we do not consider a malicious researcher. It is not possible to solve the problem we are tackling if a (malicious) researcher uses a fake dataset or strategically adds noise to the partial datasets. With the knowledge of the proposed verification algorithm, a malicious researcher can identify the conditions for which the verification algorithm has the least confidence and provide its statistics accordingly. Or, they can add the noise to the partial noisy dataset ($D_k^{\epsilon}$) in such a way that the error in the metadata may compensate the error in the computation of statistics. A verifier can not detect such cases if they do not have access to the researcher's dataset. On the other hand, in case of a fake dataset, people that are involved in the research, other researchers that work in similar studies, or verifiers that will have access to the dataset after signing data usage agreements will eventually identify that the data is fabricated. Using a fake dataset, apart from being unethical, if detected, will have serious consequences for the researcher (e.g., deteriorating the researcher's credibility among colleagues and funding agencies). Therefore, there is a huge incentive for the researcher to use a legitimate dataset. Thus, in this work, we do not consider a malicious researcher.

A researcher might (erroneously) publicly provide the correct statistics of some randomly selected SNPs rather than the ones with the strongest associations. These statistics are computed correctly, but they are not relevant as they do not belong to the SNPs with the strongest associations. Our proposed framework cannot identify such scenarios since our goal is to verify the correctness of the research findings, not their relevance. A recent work by Wang et al.~\cite{Wang2020-rc} proposed a method to verify the relevance of the outsourced GWAS results. In~\cite{Wang2020-rc}, first, the researcher generates synthetic (fake) SNPs that have strong associations with the considered phenotype. Then, they merge the synthetic SNPs with real ones, and send the dataset to a cloud server for the computation of GWAS statistics. At the end, the researcher checks if the synthetic SNPs are within the top $l$ SNPs returned by the server to verify the relevance of the returned results. We can integrate the idea of generating synthetic SNPs to our proposed framework and ensure that the verifier can check both the relevance and the correctness of the research findings.
\section{Conclusion}\label{sec:conclusion}

Providing provenance is essential for scientific research as this helps with the reproducibility of the research findings. In this paper, we have proposed a framework that can be used by a client (verifier) to efficiently verify the correctness of the computations in genome-wide association studies (GWAS) with high confidence. Furthermore, we have empirically evaluated and compared the privacy risk (in terms of the vulnerability against membership inference attacks) due to (i) the released statistics as a result of GWAS, which is accepted by many research institutions and funding agencies and (ii) the proposed framework. Notably, we have shown that the privacy risk of the dataset participants does not increase due to the additional information required by the proposed framework. As a first step towards a privacy-preserving and efficient framework to verify the correctness of GWAS statistics, we believe that this work will enable understanding the tradeoff between verification confidence of the research results and privacy leakage due to provenance. It will also help researchers to make educated decisions before publicly sharing their data.

\begin{acks}
For the research reported in this publication, Erman Ayday was partly supported by the National Library of Medicine of the National Institutes of Health under Award Number R01LM013429 and by the National Science Foundation (NSF) under grant number OAC-2112606. Xiaoqian Jiang is CPRIT Scholar in Cancer Research (RR180012), and he was supported in part by Christopher Sarofim Family Professorship, UT Stars award, UTHealth startup, the National Institute of Health (NIH) under award number R01AG066749 and U01TR002062. Jaideep Vaidya was partly supported by the National Institutes of Health award R35GM134927, and a research gift received from Cisco University Research. The content is solely the responsibility of the authors and does not necessarily represent the official views of the agencies funding the research.
\end{acks}

\section*{Availability}
Relevant code for the above experiments can be found at \\ \url{https://github.com/SpidLab/GWAS-Verification}.

\bibliographystyle{ACM-Reference-Format}
\bibliography{references}
\begin{appendices}

\section{Symbols and Notations}\label{app:notation}

Table~\ref{table:notation} summarizes the symbols and the notations used in this paper.
\begin{table}[ht!]
    \caption{Frequently used symbols and notations.}
    \begin{tabular}{|c|l|}
        \hline
        $D$ & researcher's dataset \\ \hline
        \multirow{2}{*} {$n$} & number of individuals in the GWAS study \\ & conducted by the researcher \\ \hline
        \multirow{2}{*} {$m$} & number of SNPs in the GWAS study conducted \\ & by the researcher \\ \hline
        $t$ & phenotype (trait) being studied \\ \hline
        $R$ & GWAS results returned by the researcher \\ \hline
        $l$ & number of returned SNPs by the researcher \\ \hline
        $k$ & number of SNPs in the partial noisy dataset \\ \hline
        $o$ & odds ratio \\ \hline
        $p$ & $p$-value \\ \hline
        $a$ & minor allele frequency \\ \hline
        $D_k^{\epsilon}$ & partial noisy dataset provided as part of the metadata \\ \hline
        $Q$ & GWAS results on $D_k^{\epsilon}$ \\ \hline
        $E$ & public available genomic dataset \\ \hline
        $G$ & GWAS results on E \\ \hline
        $H$ & GWAS results on $E_l^{\epsilon}$ \\ \hline
        $F$ & third dataset (the one simulating $D$) \\ \hline
        $U$ & utility loss \\ \hline
        \multirow{3}{*} {$\Phi$} & error; distance between the deviation of each statistic \\ & in $D$ (or $F$) and the deviation of that particular \\ &  statistic in $E$ \\ \hline
    \end{tabular}
	\label{table:notation}
	\vspace{-10pt}
\end{table}

\section{Quality Control Procedure}\label{app:qc_steps}
As discussed, it is crucial to follow the quality control procedure before doing GWAS on a dataset. QC steps~\cite{turner2011quality} can be divided into three main categories: (i) sample (individual) QC, which includes sex inconsistencies and chromosomal anomalies, sample relatedness, population substructure, and sample genotyping efficiency/call rate, (ii) marker (SNP) QC, which includes marker genotyping efficiency/call rate, control sample reproducibility, minor allele frequency, and Hardy-Weinberg equilibrium, and (iii) batch effects. Sex inconsistencies is one of the first steps of the QC protocol where it is checked if there is any inconsistency between the sex reported by each individual with the one predicted by the genetic data. Sample relatedness step helps to examine if there exist duplicate individuals or relatives of different degrees. By analyzing the population substructure one ensures that the study samples belong to a homogeneous population. In gentoype efficiency, it is checked if any of the individuals or SNPs should be eliminated due to providing poor quality. In control sample reproducibility, researchers remove SNPs that provide low reproducibility. Researchers also filter SNPs based on the minor allele frequency because SNPs with low MAFs (rare SNPs) have low statistical power. Via Hardy-Weinberg equilibrium, it is checked if allele and genotype frequencies remain constant over generations. Due to the high number of samples used in GWAS, samples are generally partitioned into batches. Inaccuracies during genotyping or an imbalance between the number of case and control users in each batch results in batch effect. Thus, researchers should check their data for potential batch effects. In the end, the (filtered) data obtained at the end of each category are combined together to retrieve the post-QC data.

\section{TNR Values for Different $\epsilon$ Values}\label{app:varying_eps}
\begin{figure*}[ht!]
    \centering
    \begin{subfigure}[Lactose Intolerance ($D1$)]{\includegraphics[scale=0.4]{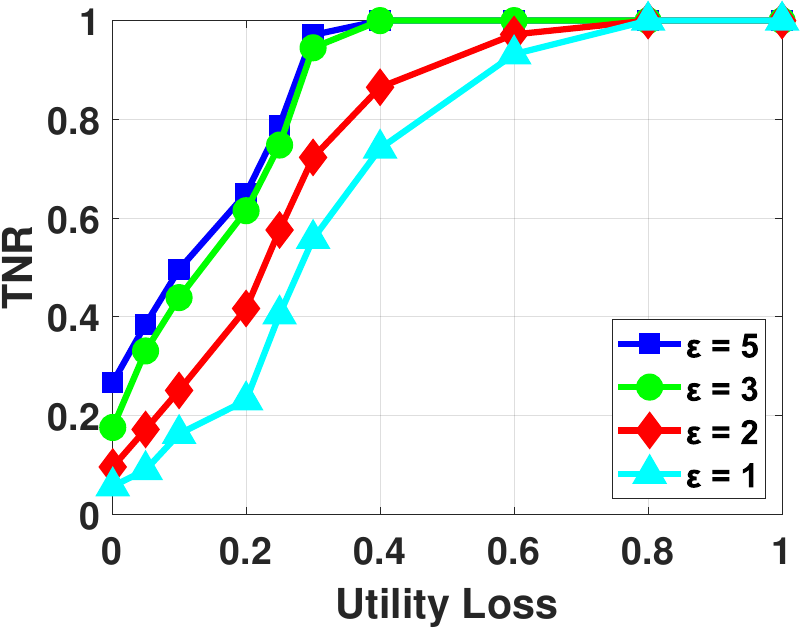}}
    \end{subfigure}\hfill
    \begin{subfigure}[Hair Color ($D2$)]{\includegraphics[scale=0.4]{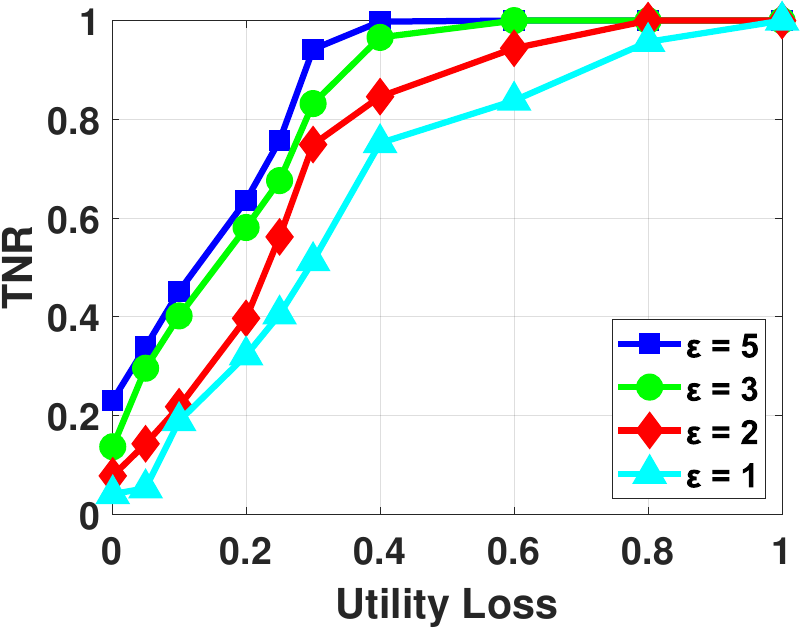}}
    \end{subfigure}\hfill
    \begin{subfigure}[Handedness ($D3$)]{\includegraphics[scale=0.4]{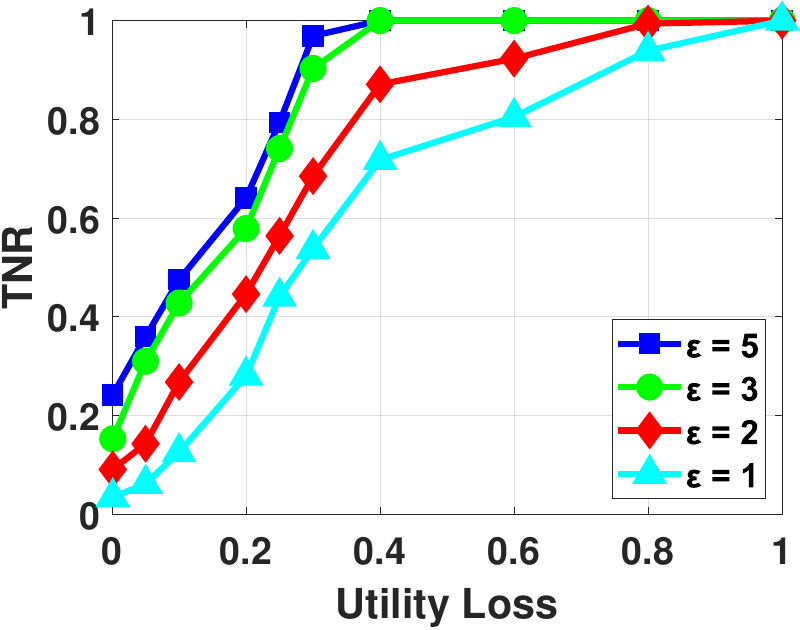}}
    \end{subfigure}\hfill
    \caption{Variation of TNR in the verification of odds ratio with respect to its utility loss for different $\epsilon$ values, when the number of released statistics is $100$ in $D1$, $D2$, and $D3$.}
    \label{fig:tnr_vs_eps_or}
    %\vspace{-10pt}
\end{figure*}

\begin{figure*}[ht!]
    \centering
    \begin{subfigure}[Lactose Intolerance ($D1$)]{\includegraphics[scale=0.4]{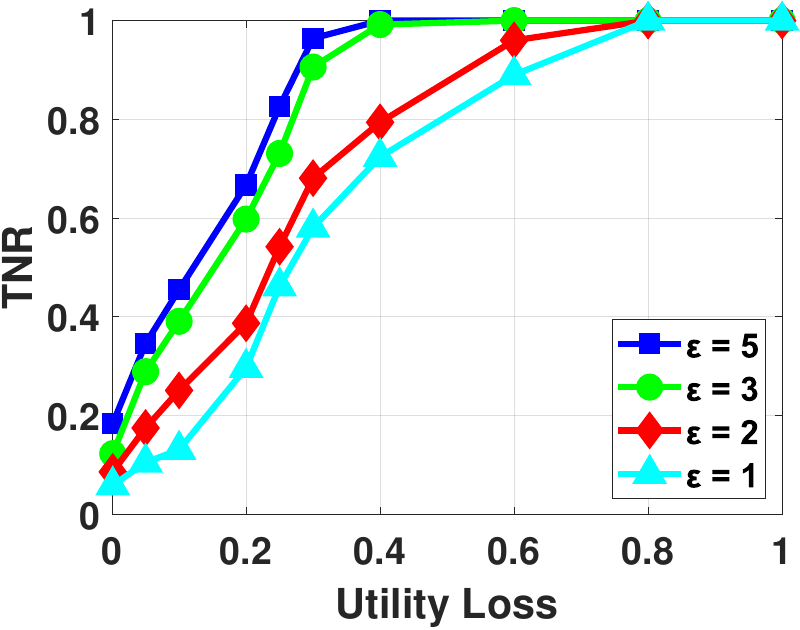}}
    \end{subfigure}\hfill
    \begin{subfigure}[Hair Color ($D2$)]{\includegraphics[scale=0.4]{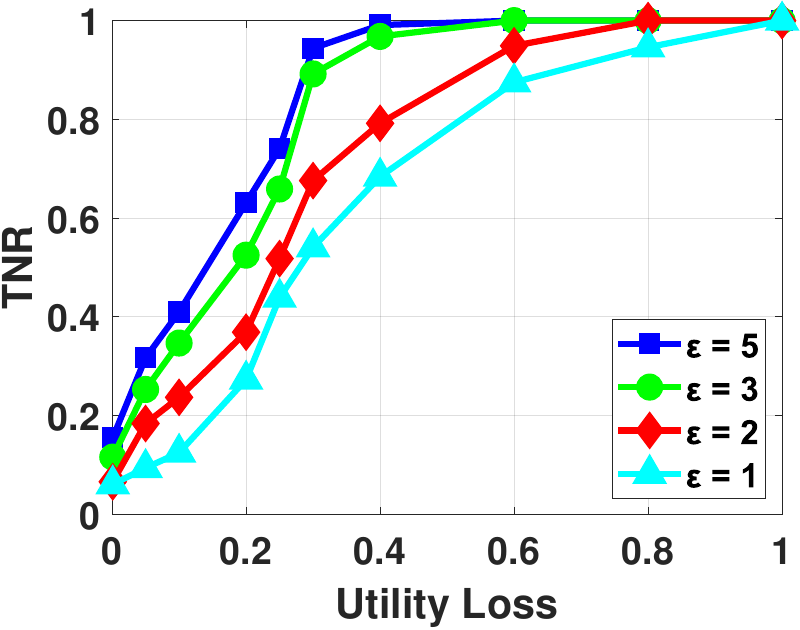}}
    \end{subfigure}\hfill
    \begin{subfigure}[Handedness ($D3$)]{\includegraphics[scale=0.4]{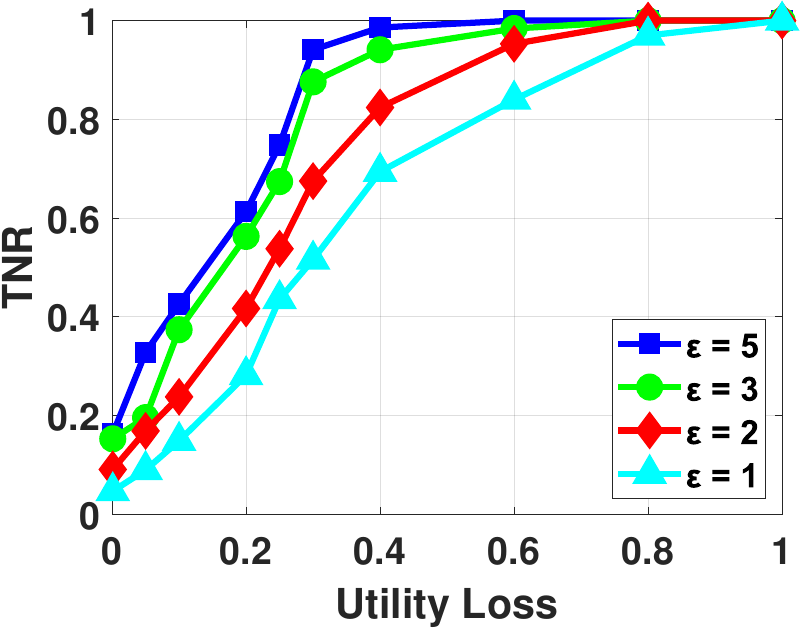}}
    \end{subfigure}\hfill
    \caption{Variation of TNR in the verification of minor allele frequency with respect to its utility loss for different $\epsilon$ values, when the number of released statistics is $100$ in $D1$, $D2$, and $D3$.}
    \label{fig:tnr_vs_eps_maf}
    %\vspace{-10pt}
\end{figure*}

Here, we show the variation of TNR with respect to its utility loss for different $\epsilon$ values and $l=100$ when verifying the correctness of odds ratio and minor allele frequencies in Figures~\ref{fig:tnr_vs_eps_or}~and~\ref{fig:tnr_vs_eps_maf}, respectively.

\section{Effect of the Number of Returned Statistics}\label{app:varying_l}

\begin{figure*}[ht!]
    \centering
    \begin{subfigure}[Lactose Intolerance ($D1$)]{\includegraphics[scale=0.4]{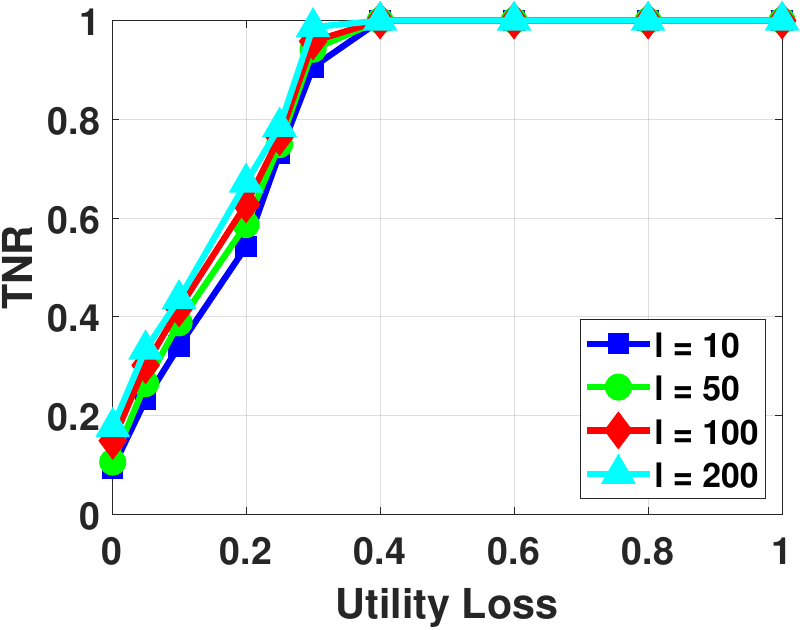}}
    \end{subfigure}\hfill
    \begin{subfigure}[Hair Color ($D2$)]{\includegraphics[scale=0.4]{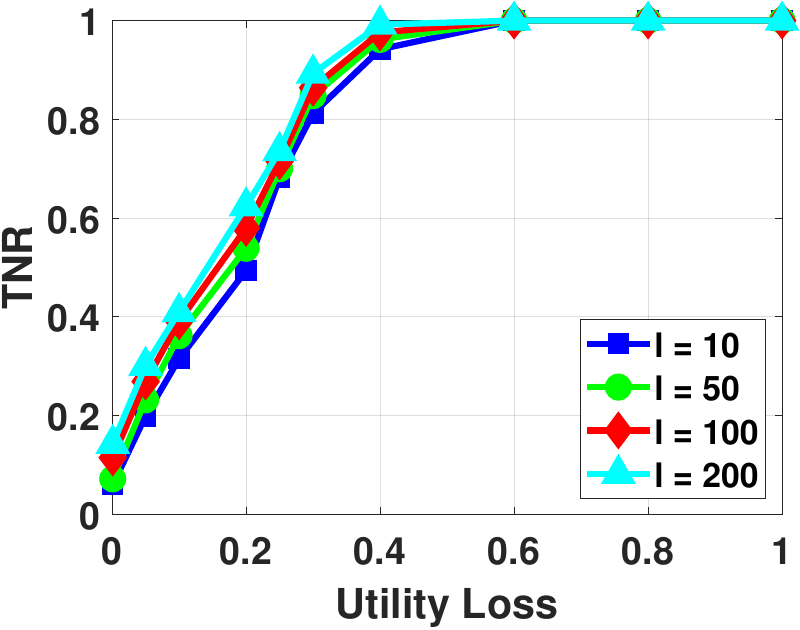}}
    \end{subfigure}\hfill
    \begin{subfigure}[Handedness ($D3$)]{\includegraphics[scale=0.4]{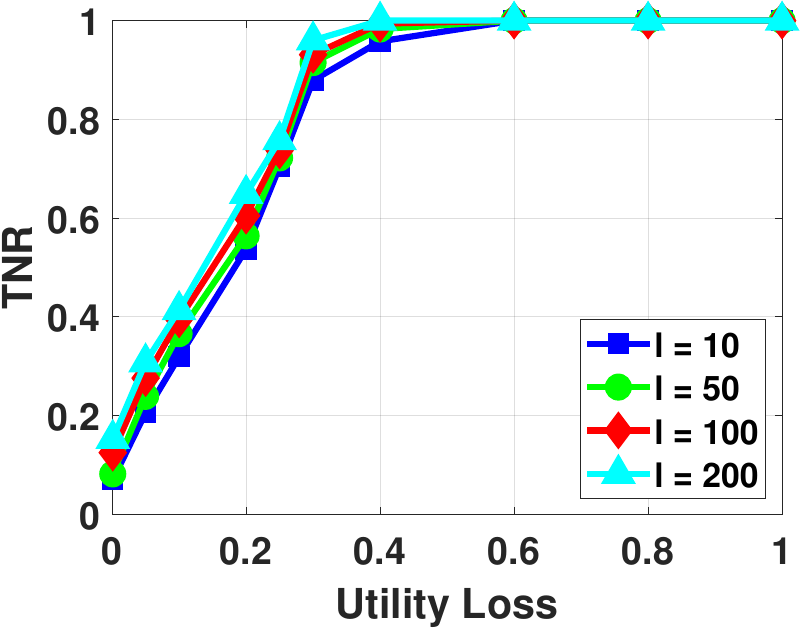}}
    \end{subfigure}\hfill
    \caption{Variation of TNR in the verification of $p$-value with respect to its utility loss for different $l$ values, when $\epsilon=5$ in $D1$, $D2$, and $D3$.}
    \label{fig:tnr_vs_l_p_value}
    %\vspace{-10pt}
\end{figure*}
We study the effect of the number of statistics returned by the researcher ($l$) to the TPR and TNR values. Table~\ref{table:tpr_vs_l} shows the TPR values achieved by the proposed framework for each statistic for $\epsilon=5$ and $l\in \{10, 50, 100, 200\}$. Figure~\ref{fig:tnr_vs_l_p_value} shows the variation of TNR with respect to the utility loss when verifying the correctness of $p$-values for the same setting. We observe that TPR and TNR slightly increase as the number of returned SNPs ($l$) increases. We obtain similar results for the other two statistics (odds ratio and minor allele frequency). Thus, we conclude that the number of statistics returned by the researcher ($l$) does not have a significant effect in the TPR and TNR values. As in previous experiments, it is harder to determine the correctness of the statistics when the returned incorrect statistical values are closer to the correct ones, but in that case, the statistics still have a high utility. 
\begin{table}[ht!]
    \centering
    \caption{TPR for verifying the correctness of $p$-value for varying number of returned SNPs ($l$ values), when $\epsilon=5$ in $D1$, $D2$, and $D3$.}
    %\vspace{-5pt}
    \begin{tabular}{|c|c|c|c|c|c|}
        \hline
        Dataset & $l=10$ & $l=50$ & $l=100$ & $l=200$ \\ \hline
        $D1$
        & $0.83$ & $0.86$ & $0.9$ & $0.93$ \\ \hline
        $D2$
        & $0.81$ & $0.86$ & $0.87$ & $0.9$ \\ \hline
        $D3$
        & $0.79$ & $84$ & $0.86$ & $0.91$ \\ \hline
    \end{tabular}
	\label{table:tpr_vs_l}
	%\vspace{-15pt}
\end{table}

\section{Adding Noise to the GWAS Statistics} 
\label{app_dp}
\begin{figure}[ht!]
    \centering
    \begin{subfigure}[Correct Statistics]{\includegraphics[scale=0.29]{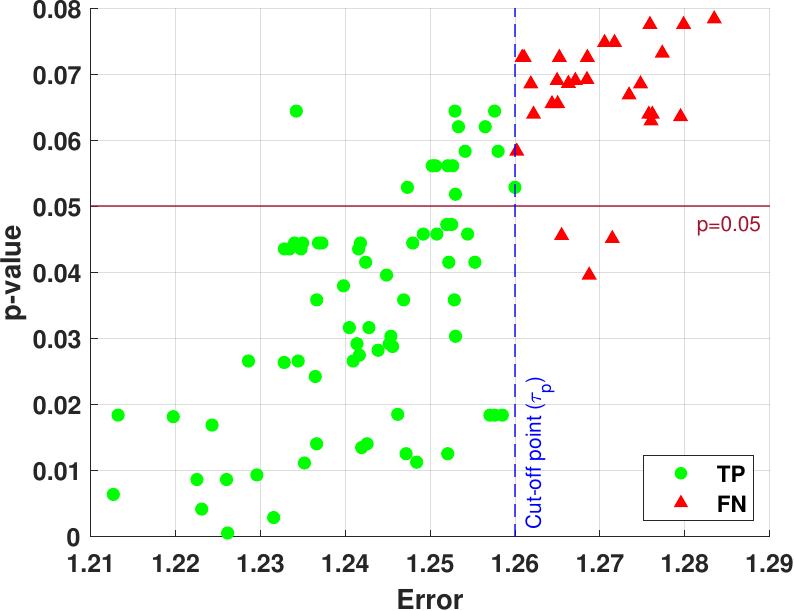}}
    \end{subfigure}\hfill
    \begin{subfigure}[Incorrect Statistics]{\includegraphics[scale=0.29]{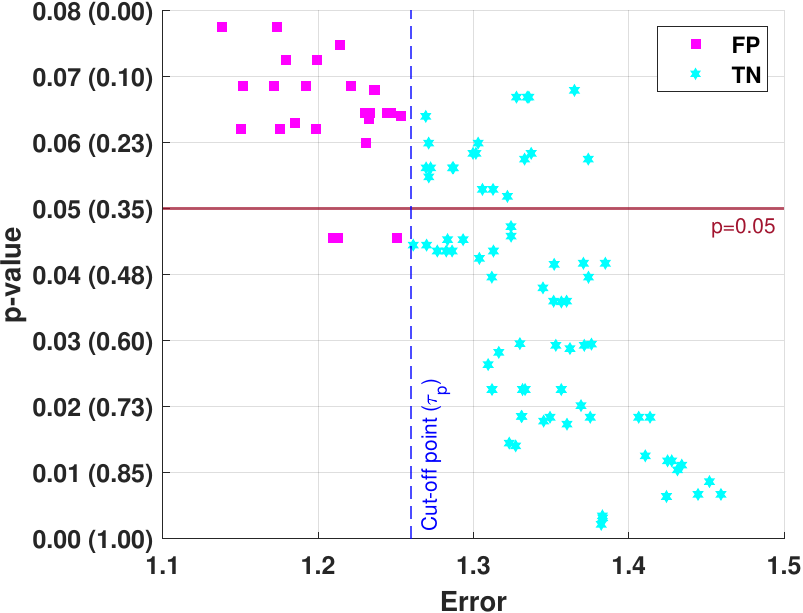}}
    \end{subfigure}\hfill
    \caption{Error of the $p$-values ($\Phi_p$) with respect to their reported $p$-values for both correct and incorrect statistics for $\epsilon=3$ and $\epsilon_{DP}=3$ in $D1$. For the incorrect statistics, the values in parenthesis (on the y-axis) show the utility loss for the corresponding reported $p$-value when its correct $p$-value is $.08$.}
    \label{fig:error_vs_p_value_dp}
\end{figure}

Here, we evaluate the performance of the proposed framework when the GWAS statistics are released under DP (with $\epsilon_{DP}=1$ or $\epsilon_{DP}=3$) and when only strong associations are considered. We select $100$ correctly computed statistics whose $p$-values are in the range $[0-.08]$ as in Figure~\ref{fig:error_vs_p_value} (Section~\ref{sec:verification}). In Figure~\ref{fig:error_vs_p_value_dp}(a), we show the $p$-values provided by the researcher after noise addition using an $\epsilon_{DP}$ of $3$ and the error ($\Phi_p$) obtained by the verifier with respect to the cut-off point ($\tau_p$). We observe that the verifier can correctly classify most of the strong associations (even when the released statistics are differentially private) achieving a TPR of $0.95$. In Figure~\ref{fig:error_vs_p_value_dp}, we also show the error of $100$ incorrect statistics whose correct $p$-values are around $.08$ (before noise addition). We observe that the verifier achieves a TNR of $0.94$ for $\epsilon_{DP}=3$ when a researcher erroneously oversells a weak association as a strong one. The TPR and TNR values achieved by the proposed framework when $\epsilon_{DP}=3$ are slightly lower than the ones achieved when no noise is added to the $p$-values (as shown in Figure~\ref{fig:error_vs_p_value}). Thus, the verifier can detect with high confidence whether the provided statistics are correct. At the same time, the researcher can ensure that the privacy of the research participants is preserved.

\begin{figure}[ht!]
    \centering
    \begin{subfigure}[Correct Statistics]{\includegraphics[scale=0.29]{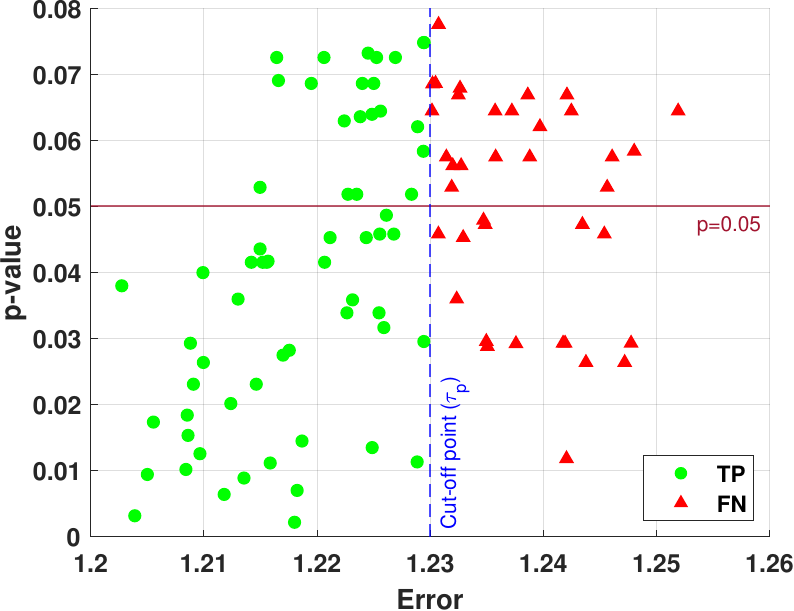}}
    \end{subfigure}\hfill
    \begin{subfigure}[Incorrect Statistics]{\includegraphics[scale=0.29]{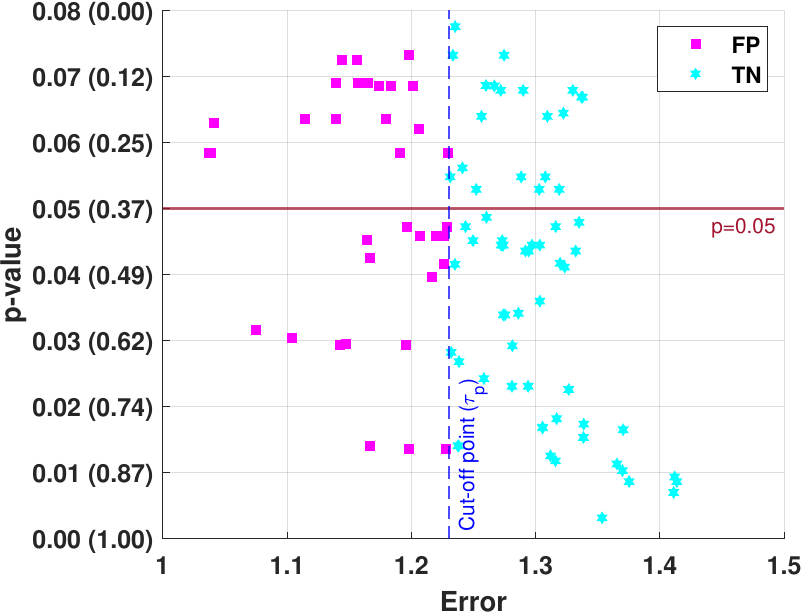}}
    \end{subfigure}\hfill
    \caption{Error of the $p$-values ($\Phi_p$) with respect to their reported $p$-values for both correct and incorrect statistics for $\epsilon=3$ and $\epsilon_{DP}=1$ in $D1$. For the incorrect statistics, the values in parenthesis (on the y-axis) show the utility loss for the corresponding reported $p$-value when its correct $p$-value is $.08$.}
    \label{fig:error_vs_p_value_dp_1}
    %\vspace{-10pt}
\end{figure}

Figure~\ref{fig:error_vs_p_value_dp_1}(a) shows the $p$-values provided by the researcher after noise addition using $\epsilon_{DP}=1$ and the error ($\Phi_p$) obtained by the verifier with respect to the cut-off point ($\tau_p$). We observe that the verifier achieves a TPR of $0.7$. Also, Figure~\ref{fig:error_vs_p_value_dp_1}(b) shows the error of $100$ incorrect statistics whose correct $p$-values are around $.08$ (before noise addition). We observe that the verifier achieves a TNR of $0.72$ for $\epsilon_{DP}=1$ when a researcher erroneously oversells a weak association as a strong one. The TPR and TNR values achieved by the proposed framework when $\epsilon_{DP}=1$ are significantly lower than the ones achieved when no noise or a smaller amount of noise ($\epsilon_{DP}=3$) is added to the $p$-values (as shown in Figures~\ref{fig:error_vs_p_value} and~\ref{fig:error_vs_p_value_dp}, respectively). However, it is worth noting that when $\epsilon_{DP}=1$, the utility of the shared statistics significantly degrades (as also shown in~\cite{johnson2013privacy,uhlerop2013privacy}), and such statistics become uninformative for other researchers.

\section{Constructing $D_k^{\epsilon}$ via Sampling} \label{sec:dataset_partioning}

\begin{figure}[ht!]
    \centering
    \includegraphics[scale=0.38]{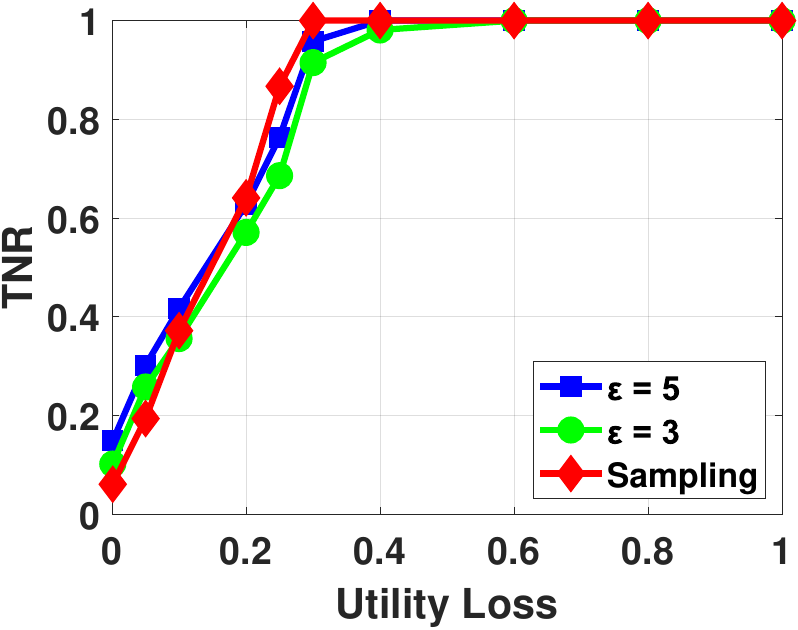}
    \caption{Variation of TNR in the verification of $p$-value with respect to its utility loss when the partial dataset is created via sampling in $D1$, for $l=100$.}
    \label{fig:sampling}
    %\vspace{-15pt}
\end{figure}

To reduce the potential privacy loss due to the partial noisy dataset, the researcher may also create the partial dataset $D_k^{\epsilon}$ by (i) first, partitioning $D$ into $b$ partitions of approximately equal size, and then (ii) randomly sampling each SNP from these $b$ partitions to construct $D_k^{\epsilon}$. This way, the number of case and control users is smaller than those in $D$, and $k$ SNPs provided for each user are sampled from the entire dataset $D$. We assume the researcher does not add noise to the partial dataset after sampling, thus $\epsilon \to \infty$. For evaluation, we use $D1$, fix $l$ to $100$, and partition $D1$ into $b=3$ parts. Figure~\ref{fig:sampling} displays the variation of TNR when verifying the correctness of $p$-value with respect to its utility loss for sampling. In the same figure, we also show the variation of TNR for $\epsilon=3$ and $\epsilon=5$ for the LDP-based technique in Section~\ref{sec:verification}. We observe that the proposed framework achieves a TPR of $0.88$ for sampling, $0.73$ for $\epsilon=3$, and $0.9$ for $\epsilon=5$, respectively. At the same time, a (misbehaving) verifier can achieve a power of only $0.08$ via the hamming distance attack (in Section~\ref{sec:edit_distance}) on the partial dataset constructed via sampling. Thus, we conclude that by using sampling, a researcher can both provide a high confidence to the verifier and preserve the privacy of the research participants. On the other hand, sampling provides more room to a researcher to do computational errors while generating the partial dataset. We leave the analysis of these (potential) errors for future work.

\end{appendices}

\end{document}